\documentclass[pdflatex,sn-nature]{sn-jnl}

\usepackage{graphicx}%
\usepackage{multirow}%
\usepackage{amsmath,amssymb,amsfonts}%
\usepackage{amsthm}%
\usepackage{mathrsfs}%
\usepackage[title]{appendix}%
\usepackage{xcolor}%
\usepackage{textcomp}%
\usepackage{manyfoot}%
\usepackage{booktabs}%
\usepackage{algorithm}%
\usepackage{algorithmicx}%
\usepackage{algpseudocode}%
\usepackage{listings}%

\newcommand{\SNote}{Supplementary Note}
\newcommand{\SFig}{Supplementary Figure}
\newcommand{\STab}{Supplementary Table}
\graphicspath{{./}{Figures/Supplementary/}}
\makeatletter
\def\SN@fmt@section{\SNote~\thesection:\hskip.5em}
\newcommand{\startsupplement}{%
  \clearpage
  \setcounter{section}{0}\setcounter{subsection}{0}%
  \setcounter{figure}{0}\setcounter{table}{0}\setcounter{equation}{0}%
  \renewcommand{\thesection}{\arabic{section}}%
  \renewcommand{\thesubsection}{\thesection\Alph{subsection}}%
  \renewcommand{\theequation}{S\arabic{equation}}%
  \renewcommand{\thefigure}{S\arabic{figure}}%
  \renewcommand{\thetable}{S\arabic{table}}%
  \renewcommand{\theHsection}{S\arabic{section}}%
  \renewcommand{\theHsubsection}{S\arabic{section}.\Alph{subsection}}%
  \renewcommand{\theHequation}{S\arabic{equation}}%
  \renewcommand{\theHfigure}{S\arabic{figure}}%
  \renewcommand{\theHtable}{S\arabic{table}}%
  \renewcommand{\figurename}{\SFig}%
  \renewcommand{\tablename}{\STab}%
  \def\@seccntformat##1{\@ifundefined{SN@fmt@##1}{\csname the##1\endcsname\hskip.5em}{\csname SN@fmt@##1\endcsname}}%
  \let\SN@lsection\l@section
  \renewcommand*\l@section[2]{\begingroup\def\numberline####1{\SNote~####1:\ }\SN@lsection{##1}{##2}\endgroup}%
  \setlength{\parindent}{0pt}\setlength{\parskip}{0.5em}%
  \setlength{\emergencystretch}{2em}%
}
\makeatother

\begin{document}
\addtocontents{toc}{\protect\setcounter{tocdepth}{-10}}

\title[Training wavefronts through disorder]{Single-shot holographic gradients for training wavefronts through disorder}

\author[1]{\fnm{Kerem} \sur{Ekec}}
\equalcont{These authors contributed equally to this work.}

\author[2]{\fnm{Bahad{\i}r Utku} \sur{Kesgin}}
\equalcont{These authors contributed equally to this work.}

\author*[1,2]{\fnm{U\u{g}ur} \sur{Te\u{g}in}}\email{utegin@ku.edu.tr}

\affil[1]{\orgdiv{Department of Biomedical Sciences and Engineering}, \orgname{Ko\c{c} University}, \orgaddress{\city{\.{I}stanbul}, \postcode{34450}, \country{T\"{u}rkiye}}}

\affil[2]{\orgdiv{Department of Electrical and Electronics Engineering}, \orgname{Ko\c{c} University}, \orgaddress{\city{\.{I}stanbul}, \postcode{34450}, \country{T\"{u}rkiye}}}

\abstract{Training optical systems in situ needs the gradient of their loss, yet a camera records only intensity and the medium is never exactly known. We show that one off-axis hologram gives this gradient, an approach we call holographic gradients. Light passes through the medium only forward; the hologram measures the error, and a compact adjoint, fitted to ten probes, is refreshed in seconds during training. Each step costs one modulated pattern. Through a thin diffuser, it focuses light and forms images as well as a measured transmission matrix, without its thousands of patterns. Under controlled drift, refreshing the adjoint recovers a grating to 17.6~dB, against 9.1~dB for a frozen model and 19.5~dB without drift, and training succeeds through a thick diffuser. Trained offline, the reference beam becomes a computing layer that plays a video game. Holographic gradients let optical systems learn while they operate, from imaging through dynamic media to self-correcting processors.}


\maketitle

\label{sec:intro}

Photonics and artificial intelligence (AI) have a mutualistic relationship that runs in two directions. In one direction, wavefront shaping helps light pass through various types of complex media~\cite{mosk2012controlling,rotter2017light}. Classical approaches optimize the wavefront iteratively with the measured intensity as feedback~\cite{vellekoop2007focusing}, measure the transmission matrix (TM) of the medium~\cite{popoff2010measuring,yu2013measuring}, or record the scattered field and phase conjugate it~\cite{yaqoob2008optical}, and AI increasingly learns these corrections from data. In the other direction, light performs the computations of artificial intelligence~\cite{wetzstein2020inference,mcmahon2023physics}. Optical neural networks harness the parallelism of light in free space~\cite{lin2018all,hamerly2019large}, integrated circuits~\cite{shen2017deep}, scattering media~\cite{carpinlioglu2025genetically,rafayelyan2020large,saade2016random} and multimode fibers~\cite{tegin2021scalable,oguz2025training,kesgin2025photonic,eslik2026recurrent}. Both of these directions share a common problem of training a physical system that is never exactly known and whose gradient does not come for free. To tackle this, various in silico and in situ methods are used for efficient and accurate training.

Optical processors are commonly designed in silico and then deployed, where fabrication errors, misalignment and drift degrade the experimental performance of the trained processor~\cite{wright2022deep,momeni2025training}. Data-driven methods try to close this gap by training neural networks to reconstruct images from scattered and lensless measurements~\cite{li2018deep,ekec2026resolution}, whereas in situ methods keep the physical system in the loop. The simplest in situ methods are model-free and treat the system as a black box, updating its parameters from measured losses with reinforcement learning~\cite{li2026model}, genetic algorithms~\cite{liu2026focusing,carpinlioglu2025genetically}, Bayesian optimization~\cite{kesgin2025fiber} or forward-only learning rules~\cite{momeni2023backpropagation,oguz2023forward}. These methods need no model, but their measurements grow with the number of parameters, often to thousands of frames per task~\cite{li2026model}. Using the known physics inside the photonic system can increase the sample efficiency of these methods.

Physics-informed methods use the gradient instead. Physics-aware training executes the forward pass on the physical system and the backward pass on a differentiable digital model learned from input and output pairs~\cite{wright2022deep}. Similar digital twins compute through nonlinear and spiking optical systems~\cite{oguz2025training,kesgin2026optical} and, learned from ten thousand or more intensity measurements, through scattering media~\cite{wang2026tron,li2018imaging}, and physical models have been fitted to measured transmission matrices~\cite{matthes2021learning}. Other approaches obtain the backward pass physically, by propagating the error back through the optical system~\cite{hughes2018training,pai2023experimentally,spall2025training}, or acquire the gradient optically to shape wavefronts through scattering media in reflection~\cite{monin2026rapid}. In these methods, the backward pass sets the cost of training because a learned model needs many training pairs and must be relearned when the system changes. To tackle this, the backward pass can be moved to the physical domain, which requires reciprocal propagation.

A camera records intensity, whereas, owing to the complex nature of light, the gradient of most optical tasks depends on the complex field. Speckle correlations and the memory effect allow non-invasive imaging through opaque layers~\cite{freund1988memory,bertolotti2012non}, interferometric measurements give the complex TM of a medium~\cite{popoff2010measuring,yu2013measuring}, model-based wavefront shaping computes corrections from a model of the sample~\cite{thendiyammal2020model}, and a few measured fields combined with a phase-screen model can replace the TM for computational correction~\cite{haim2025image}. Without a reference, the field is recovered with phase retrieval, at the cost of more measurements and a non-convex inversion~\cite{fienup1982phase,dremeau2015reference}. Recently, complex measurements have also entered physical training. Fully forward mode training measures fields with phase-shifting holography and runs the error propagation forward by exploiting spatial symmetry and Lorentz reciprocity~\cite{xue2024fully}. Physical backward passes based on time reversal and phase conjugation have trained photonic processors in situ and threaded light through dynamic complex media~\cite{zhou2026photonic,mididoddi2025threading}. These methods carry the backward pass in light, which relies on reciprocal or symmetric paths aligned in both directions and on more than one frame per training step~\cite{xue2024fully,zhou2026photonic,mididoddi2025threading}.

Off-axis holography is a simple and efficient way to access the complex field. A tilted reference beam shifts the interference term to a spatial carrier, so that one camera frame gives the complex field after Fourier filtering~\cite{leith1962reconstructed,cuche2000spatial}. In optical processors, the reference beam has served as a ruler for phase, although homodyne detection has shown that the interference with a reference can itself compute an optical inner product~\cite{hamerly2019large,chen2023deep}. Whether one off-axis frame can replace the backward pass, so that light only runs forward and the reference beam is itself trained for optical processing, has not been tested.

Here, we show that a single off-axis hologram gives the gradient of an optical system behind a scattering medium. We call this framework holographic gradients. Light passes through the medium only forward; the hologram measures the error, and a compact adjoint is refreshed in seconds during training. The bench executes the forward pass, and one camera frame gives the measured complex output field and therefore the error field of the loss. Only the backward step is computed, through a physics model of the medium fitted from measured probe fields. Because the error is measured, a coarse model suffices. As the probe fields are complex, the model is linear in the medium, and its fit needs tens to hundreds of probes rather than a full TM or a learned twin. Since the error is never sent back into the optics, no reciprocal path or second optical pass has to be aligned. Each step costs one SLM pattern, and the same frames refit the model while the medium drifts and extend it to thick diffusers.

In the off-axis geometry, two arms of the system are trainable. Data is carried by the object arm with a trainable phase, and the reference arm bypasses the medium with a trainable complex field. Since the two arms meet only at the camera, the same measured error field gives the gradients of both. The object arm is trained with the medium in the loop. On the other hand, the reference arm treats the medium as a fixed optical layer where the fields of a dataset are measured once, and the reference is trained offline as an optical readout. Due to the different tilts of the two arms, they can be demodulated and trained together in situ. Thus, one setup covers medium-based tasks, such as imaging through disorder, and task-based computing on fields that the medium has already processed.

We show that holographic gradients adapt across tasks, apply to both arms of the hologram and remain effective when the medium changes, through focusing, imaging and classification behind a diffuser, game control with an offline-trained reference, and training through drifting and thick diffusers. Through a ground-glass diffuser, the framework matches a measured TM in focusing and imaging from more than an order of magnitude fewer SLM patterns, and the offline-trained reference wins every episode of ViZDoom Basic in closed loop, like an oracle that reads the game state. While the diffuser drifts, refitting the model from a few frames keeps the focus that a frozen model loses, and a two-screen model fitted from 200 probe fields trains light through a thick diffuser whose memory effect has collapsed. Together, these results show that one hologram per step trains an optical system through drifting and thick disorder.

\section{Results}\label{sec:results}

\subsection*{Obtaining holographic gradients}

\begin{figure}[!htb]
\centering
\includegraphics[width=\linewidth]{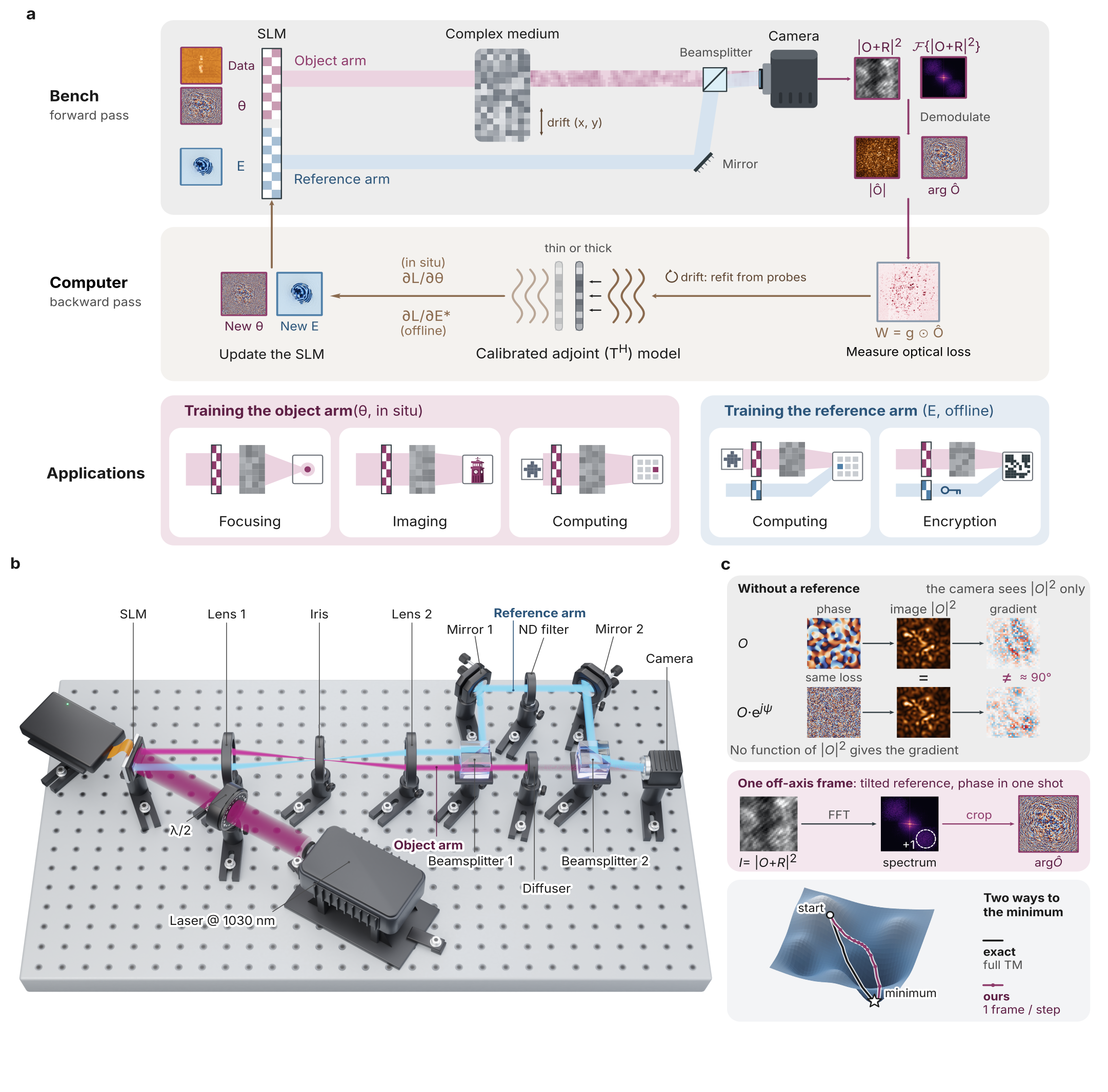}
\caption{\textbf{Holographic gradients.} \textbf{a}, The bench executes the forward pass; one off-axis frame gives the measured field $\hat O$ and the error field $W$, and a fitted adjoint carries $W$ back to the object arm $\theta$ (in situ) or the reference arm $E$ (offline). Bottom, demonstrated tasks. \textbf{b}, Experimental setup. \textbf{c}, Why a reference is needed, and the off-axis readout.}\label{fig1}
\end{figure}

Figure~\ref{fig1} illustrates how one hologram gives both the output field and the gradient. A phase-only spatial light modulator (SLM) is divided into two regions (Fig.~\ref{fig1}b). The object region carries the input data and a trainable phase pattern $\theta$, and its light passes through a diffuser, whereas the reference region carries a trainable complex pattern $E$ that bypasses the diffuser and meets the object beam on the camera at a small angle. The camera records
\begin{equation}
X_O = x\odot e^{j\theta},\qquad O = T_O\{X_O\},\qquad R = T_R\{E\},\qquad I = |O+R|^{2},
\label{eq:arch}
\end{equation}
where $x=e^{j\chi}$ is the data written as a phase, and the linear operators $T_O$ and $T_R$ carry the object and reference arms to the camera fields $O$ and $R$. Since the reference is tilted, the interference term $OR^{*}$ sits on a spatial carrier, and Fourier filtering gives the measured complex field $\hat O$ from a single frame~\cite{cuche2000spatial} (Supplementary Note~1B, Supplementary Fig.~S1).

The same frame also gives the error of the loss. We differentiate a real loss $L$ computed from the camera with respect to the complex fields by Wirtinger calculus~\cite{wirtinger1927formalen}. For a loss on the measured object, the error field is
\begin{equation}
W_O = \frac{\partial L}{\partial O^{*}} = g\odot\hat O,\qquad g=\frac{\partial L}{\partial |\hat O|^{2}},
\label{eq:errorfield}
\end{equation}
and since the arms meet only at the camera, the error field $W_R=\partial L/\partial R^{*}$ of the reference arm is likewise formed from $\hat O$. The chain rule gives the gradients of both arms,
\begin{equation}
\frac{\partial L}{\partial \theta} = 2\,\mathrm{Im}\big[X_O^{*}\odot T_O^{\mathrm{H}}\{W_O\}\big],\qquad
\frac{\partial L}{\partial E^{*}} = T_R^{\mathrm{H}}\{W_R\},
\label{eq:gradients}
\end{equation}
where the adjoint $^{\mathrm{H}}$ carries the error field from the camera back to the SLM (Supplementary Note~2A). Here, the error field is measured, up to one phase per frame taken from the model, and only the adjoints are modelled. The object-arm gradient requires a model of the diffuser, whereas the reference-arm gradient requires only the free-space reference path of the reference and is independent of the object arm, even a nonlinear one (Supplementary Note~2A). 

The phase carried by the reference is essential to this gradient (Fig.~\ref{fig1}c). The intensity $|O|^{2}$ does not change when each pixel of $O$ acquires a phase, whereas the backward field $T_O^{\mathrm H}\{g\odot O\}$ does, so two fields with the same image can have nearly orthogonal gradients (Supplementary Note~2B). Accordingly, a model trained on intensities alone predicts the bench poorly, and training through it forms a grating at a peak signal-to-noise ratio (PSNR) of 8.9 dB instead of 16.1~dB in the last iteration (Supplementary Fig.~S2).

\subsection*{In situ training of the object arm}

Having obtained the error field from the camera, training the object arm only requires an adjoint of the diffuser. Rather than measuring a TM, which needs one probe per SLM cell, we describe the ground-glass diffuser as a thin scattering screen~\cite{li2018imaging,haskel2018modeling,haim2025image},
\begin{equation}
T_O \underset{\text{thin}}{\cong} P_{z_s}\, \mathrm{diag}(D)\, P_{-z_s}\, F ,\qquad T_O^{\mathrm{H}} \underset{\text{thin}}{\cong} F^{\mathrm{H}}\, P_{z_s}\, \mathrm{diag}(D^{*})\, P_{-z_s} ,
\label{eq:model}
\end{equation}
where $F$ is the fixed free-space path, $P_{\pm z_s}$ propagates to and from the screen, and $D$ is the unknown screen transmission (Supplementary Note~3A). We fit the screen to $K$ measured probe fields $\hat O_k$ written as patterns $X_k$,
\begin{equation}
\hat D = \arg\min_{D} \sum_{k=1}^{K} \big\| \hat O_k - T_O[D]\{X_k\} \big\|^{2}.
\label{eq:fit}
\end{equation}
The measured field is linear in $D$, so a few probes determine the screen, although the fit is not convex (Supplementary Note~3A).

\begin{figure}[!htb]
\centering
\includegraphics[width=\linewidth]{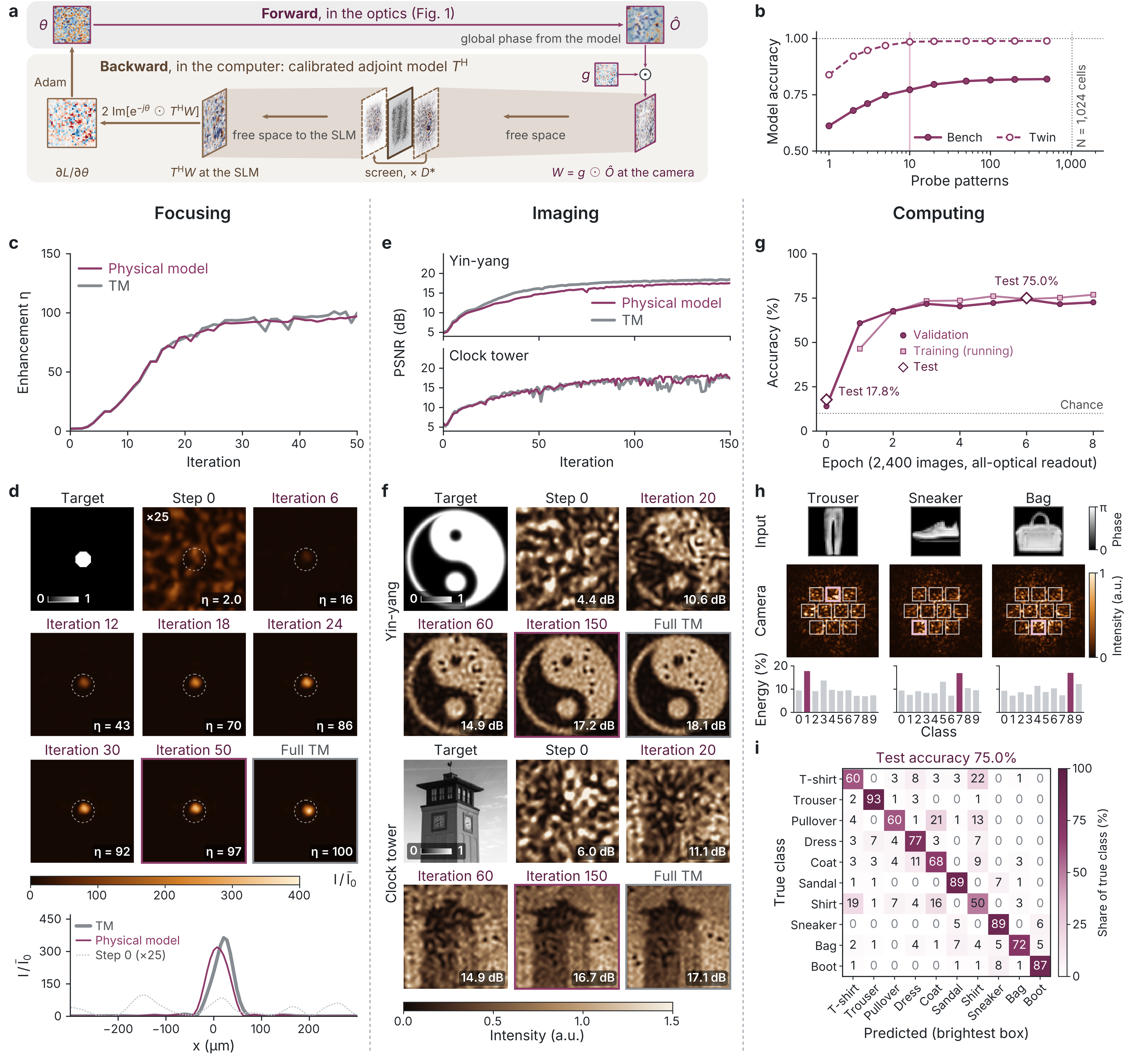}
\caption{\textbf{In situ training of the object arm.} \textbf{a}, Training step. \textbf{b}, Model accuracy versus number of probes. \textbf{c},\textbf{d}, Focusing with the physical model and the measured TM. \textbf{e},\textbf{f}, Image formation. \textbf{g}--\textbf{i}, All-optical FashionMNIST classification.}\label{fig2}
\end{figure}

Figure~\ref{fig2}a shows the resulting training step. Ten random probes are enough to fit the screen. The fitted model predicts held-out fields with a model accuracy of 0.77 on the bench and 0.985 in the digital twin, close to the plateaus of 0.82 and 0.989 reached with 500 probes (Fig.~\ref{fig2}b).We first evaluate the framework on focusing light on a single speckle grain behind the diffuser~\cite{vellekoop2007focusing}, a benchmark against the full TM. The enhancement reaches 97, on par with 100 for the TM (Fig.~\ref{fig2}c,d); both loops reach the same value, so the model does not limit the focus. Moving from a single focus to images, we form a yin-yang and a clock tower at 17.2 and 16.7~dB PSNR, on par with the TM (18.1 and 17.1~dB), from ten probes per session instead of the 3,072 patterns of the TM (Fig.~\ref{fig2}e,f; Supplementary Note~3C, Supplementary Table~S1). Over the imaging window (Fig.~\ref{fig2}b is scored over the whole field), the TM predicts held-out fields with a higher model accuracy than the physical model (0.95 against 0.90; Supplementary Note~3C). The measured error field thus matters more than the accuracy of the model.

Beyond shaping light, the object arm can also compute. We train it as an all-optical classifier of FashionMNIST images~\cite{xiao2017fashion}, and the brightest of ten detector regions of the measured field gives the class without digital weights. Trained in situ, this single phase layer raises the test accuracy from 17.8\% to 75.0\% (Fig.~\ref{fig2}g--i; Supplementary Note~3B).

\subsection*{The reference beam as a trainable layer}

\begin{figure}[!htb]
\centering
\includegraphics[width=\linewidth]{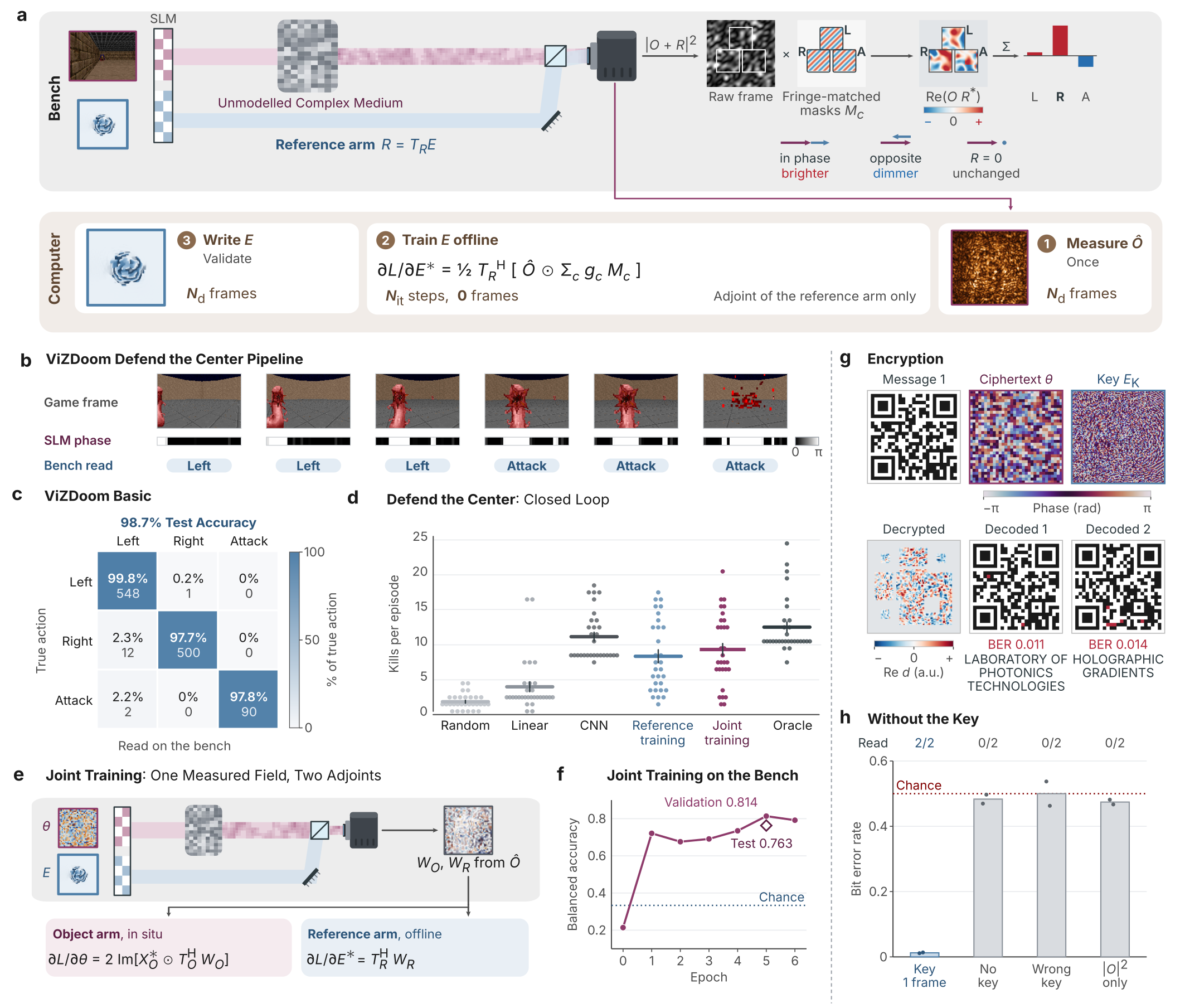}
\caption{\textbf{The reference beam as a trainable layer.} \textbf{a}, Offline training and fringe-matched readout. \textbf{b}, ViZDoom played by the bench. \textbf{c}, Basic scenario, test confusion matrix (demodulated read). \textbf{d}, Defend the Center, kills per episode (mean $\pm$ s.e.m.; all-optical read). \textbf{e},\textbf{f}, Joint training of both arms. \textbf{g},\textbf{h}, Two-key decryption of QR codes and bit error rates.}\label{fig3}
\end{figure}

The same hologram also trains the reference beam, whose adjoint is only its free-space path, so the medium needs no model. Since $O$ does not depend on $E$, the camera image of input $n$ for any reference is
\begin{equation}
I^{(n)}(E) = \big|\hat O^{(n)} + R(E)\big|^{2},
\label{eq:oneshot}
\end{equation}
where $\hat O^{(n)}$ is measured once. The medium therefore acts as a fixed optical layer, and the reference is trained offline on the stored fields (Fig.~\ref{fig3}a; Supplementary Note~4A).

To read the result optically, we place fixed fringe-matched masks on the raw camera frame, each a detector window $M_c$ multiplied by the carrier $\mathbf{f}_0$ of the off-axis fringes. These filters are bandpass filters to get the carrier from intensity without Fourier transform with carrier locations known from the reference tilt, and the score of class $c$ is
\begin{equation}
s_c = \mathrm{Re}\Big[e^{-j\phi}\sum_{\mathbf{x}} M_c(\mathbf{x})\,e^{j2\pi\mathbf{f}_0\cdot\mathbf{x}}\,I(\mathbf{x})\Big] \cong \mathrm{Re}\Big[e^{-j\phi}\sum_{\mathbf{x}\in M_c} O(\mathbf{x})\,R^{*}(\mathbf{x})\Big],
\label{eq:readout}
\end{equation}
where $\phi$ is the phase between the arms. The mask moves only $OR^{*}$ to zero frequency, whereas $|O|^{2}$, $|R|^{2}$ and $O^{*}R$ oscillate under it and cancel, so each score sums the interference term over its window without demodulation (Methods). The inner product of the diffused input with the trained reference is thus computed by interference~\cite{hamerly2019large,chen2023deep}, and the masks emulate fixed fringe-matched detectors. The reference error field is then $W_R=\partial L/\partial R^{*}=\tfrac12\,\hat O\odot\sum_c g_cM_c$, with $g_c=\partial L/\partial s_c$. Each score is linear in $E$, so training the reference is convex (Supplementary Note~2A).

We apply the offline-trained reference to game control in ViZDoom~\cite{kempka2016vizdoom}, where the bench maps each game frame to an action and plays in closed loop (Fig.~\ref{fig3}b). In the Basic scenario, the scores are read by demodulation instead of the masks. The interference term is taken from the demodulated field, and the reference is trained to maximize the classification of these scores. At the cost of one Fourier filtering per frame, this read classifies test frames with 98.7\% accuracy, close to the 99.9\% of a digital convolutional network (Fig.~\ref{fig3}c), against 91.8\% for a plain all-optical read in a separate run (Supplementary Note~4B). In closed loop, the optical actor wins all 100 unseen episodes, about one step per episode slower than an oracle that reads the game state. In the harder Defend the Center scenario, the optical reference read all-optically with the masks kills $7.9\pm0.9$ enemies per episode, more than twice a linear classifier on the input (3.5), against 10.7 for a digital convolutional network and 12.0 for the oracle (Fig.~\ref{fig3}d). On the same test frames, the masks and demodulation choose the same action for 98\% of frames (Supplementary Note~4B, Supplementary Fig.~S3). 

Beyond classification, the diffuser and the offline-trained reference act as the two keys of an optical cipher~\cite{refregier1995optical,horstmeyer2013physical}. The key is a reference pattern trained offline, and the ciphertext, a phase pattern on the object arm, is optimized in situ through the ten-probe model of the diffuser. With both keys, the bench decrypts each QR code~\cite{barrera2013optical} from one frame with a bit error rate of at most 0.014 and every code is read, whereas reads with a flat reference, a wrong key or the intensity alone stay at chance and are never read (Fig.~\ref{fig3}g,h; Supplementary Note~4C, Supplementary Fig.~S4).

\subsection*{Joint training and drift robustness}

Having trained each arm separately, we next train both together. Since both error fields come from the same measured field, $\theta$ and $E$ are updated from the same frames (Fig.~\ref{fig3}e). On Defend the Center, each epoch trains $\theta$ in situ through the ten-probe model of the diffuser, refitted on the fields of each epoch, and $E$ offline on the fields just measured. The jointly trained bench classifies test frames with a balanced accuracy of 0.76 from the raw frame and plays in closed loop with $8.8\pm0.9$ kills per episode, against 1.4 for random actions (Fig.~\ref{fig3}d,f; Supplementary Note~4D, Supplementary Fig.~S5).

\begin{figure}[!htb]
\centering
\includegraphics[width=\linewidth]{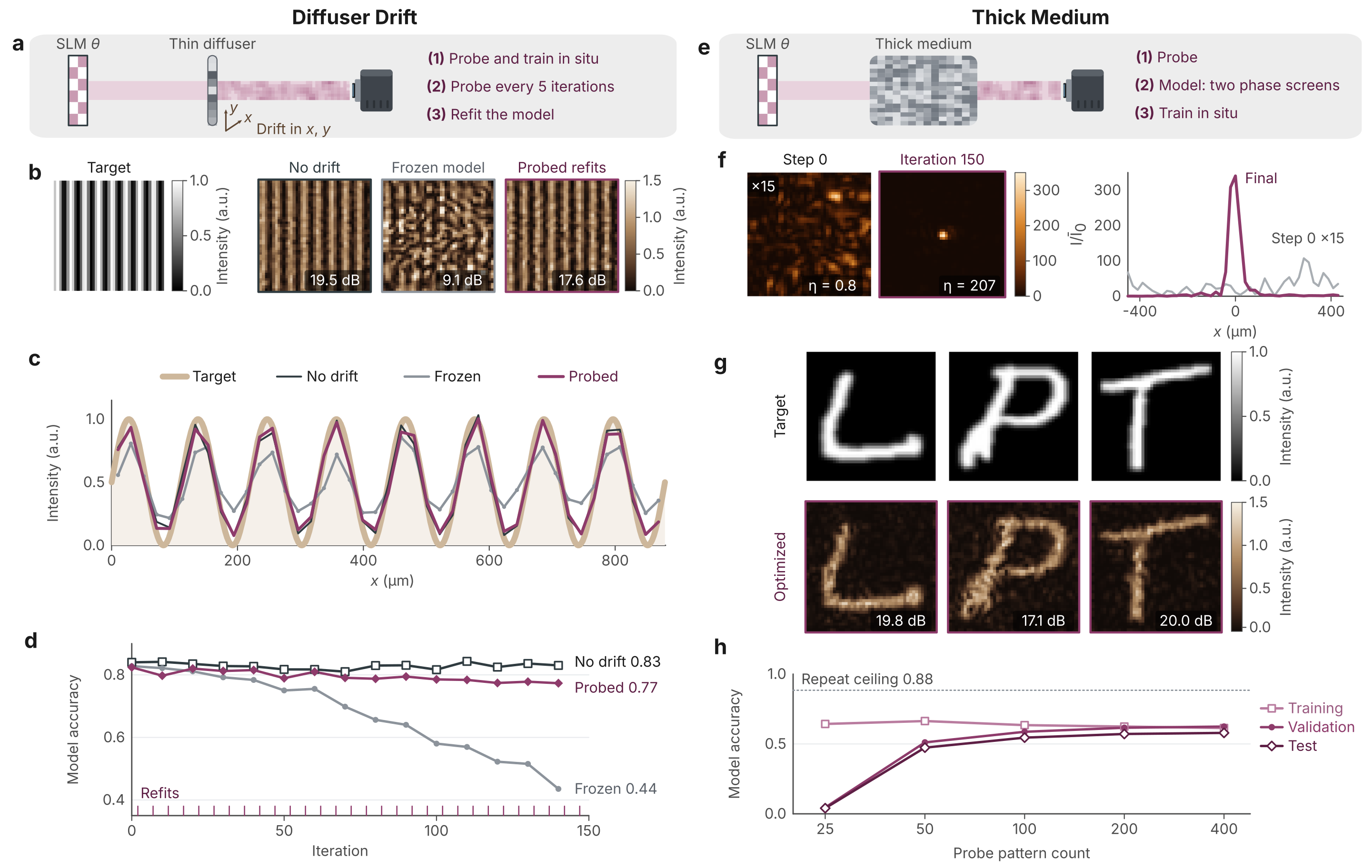}
\caption{\textbf{Drift and thick diffusers.} \textbf{a}, The diffuser drifts during training, and probes with the latest training frame refit the model. \textbf{b},\textbf{c}, Grating formed under drift and its fringe profiles without drift and with a frozen or refitted model. \textbf{d}, Model accuracy during training; ticks, refits. \textbf{e}, Thick diffuser modelled by two phase screens. \textbf{f}, Focusing through the thick diffuser; $\eta$, enhancement. \textbf{g}, EMNIST letters formed through it. \textbf{h}, Two-screen model accuracy versus number of probes; line, repeat ceiling.}\label{fig4}
\end{figure}

So far, the medium was static. Since the field is measured at every step, the forward pass and the error field remain correct when the medium drifts, and only the modelled adjoint ages. We therefore keep the adjoint current by repeating the fit of equation~(\ref{eq:fit}) during training, from the probes and training frames of the last few iterations (Fig.~\ref{fig4}a).

To test this, during training, we translate the illumination across the diffuser with a growing SLM phase ramp unknown to the model, until the medium seen by the beam decorrelates to 0.5 (Supplementary Note~5A, Supplementary Fig.~S6). 
The measured error field tolerates a coarse model, but not a stale one. With a frozen model, the model correlation decays from 0.83 to 0.44, and a grating formed in situ falls from 19.5 to 9.1~dB (PSNR of the mean image) as its fringes wash out (Fig.~\ref{fig4}b--d). Refitting every five steps from two probes and the latest training frame keeps the model near its drift-free accuracy and recovers the grating to 17.6~dB. Refits likewise keep a focus close to its drift-free enhancement, whereas a frozen model loses most of it (Supplementary Note~5A).

\subsection*{Scaling of the architecture}

The thin-screen model relies on the memory effect, which shortens as the diffuser becomes thicker~\cite{freund1988memory,osnabrugge2017generalized}. We therefore use a thick diffuser whose output speckle decorrelates under a small input tilt, so that it no longer acts as one surface (Supplementary Note~5B, Supplementary Fig.~S7). We model it with two thin screens separated by free-space propagation, a multi-slice description~\cite{kamilov2015learning} with which two separated thin diffusers have emulated a volume diffuser of narrow memory effect~\cite{lib2020real},
\begin{equation}
T_O \underset{\text{thick}}{\cong} P_{z_2}\, \mathrm{diag}(D_2)\, P_{\Delta z}\, \mathrm{diag}(D_1)\, P_{-(z_2+\Delta z)}\, F ,
\label{eq:thick}
\end{equation}
which reduces to the thin screen of equation~(\ref{eq:model}) when the separation $\Delta z$ vanishes. Both screens are fitted jointly with equation~(\ref{eq:fit}) (Fig.~\ref{fig4}e; Supplementary Note~5B).

Training through this coarse two-screen adjoint focuses light through the thick diffuser with an enhancement of 207 and forms EMNIST letters at 17 to 20~dB PSNR (Fig.~\ref{fig4}f,g). In one session, two screens focus to 228 and 194, and a single screen to only 12 and 14 (Supplementary Fig.~S7d).
Taken together with the refits under drift, these results extend holographic gradients beyond thin and static scattering surfaces. 

Besides thick media and robustness to drift, the approach can scale with the size of the system. In a digital twin of the setup with the bench noise, the ten-probe model keeps a held-out correlation of 0.84--0.87 from $N=16^{2}$ to $128^{2}$ SLM cells, whereas a TM needs at least $N$ patterns, trains to a lower enhancement than the model when limited to that minimum and surpasses it by a factor of 1.2--1.5 only from $3N$ patterns, so the cost of the adjoint stays fixed while that of a TM grows linearly with $N$ (Supplementary Note~6).

\section{Discussion}\label{sec:discussion}

Our results rest on one principle, to measure the forward field and model only the adjoint. Every off-axis frame yields the complex field, so the forward pass is always performed by the optics and the error that drives each update is measured rather than predicted. A model enters only where this error is carried back to the SLM; it need not reproduce the system exactly, only point the update downhill.

This principle has two consequences. First, a compact physical model learned from about ten probe patterns trains the hardware to the quality obtained with a fully measured TM, from more than an order of magnitude fewer SLM patterns. The TM is the most complete linear description our measurements allow and hence the practical ceiling for any model-based adjoint on this setup. Errors in the model only tilt each step; the next measurement reveals their effect, and the loop compensates. Both adjoints reach the same enhancement in the same number of steps and give the same single-step focus from their own matrix (150 and 153), and the gradient of a model fitted to a single probe already points within 45 degrees of the TM gradient (Supplementary Note~3C). Second, the model can be refreshed while the optimization runs. It is linear in the diffuser screen and fitted with exact gradients, so probing and refitting take seconds and can be interleaved with training. Under diffuser drift, a frozen model loses most of its benefit, whereas one refreshed with a few probe patterns recovers most of the focus and image quality. Calibration thus stops being a separate step performed before the experiment and becomes part of the training loop itself.

The reference arm follows the same principle in its simplest form, since it bypasses the medium and its adjoint needs only a free-space model of its own path. Since the object field does not depend on the reference pattern, the reference gradient is independent of the object arm. The diffuser then acts as a fixed random mixing layer and the trained reference as a complex weight layer, and the interference term evaluates their inner product optically in a single camera frame. This suffices to play ViZDoom in closed loop from optical decisions and to encrypt messages that are readable only with both the physical diffuser and the designed reference. The reference arm, usually a passive part of a holographic measurement, thereby becomes a trainable part of the computation, and it can be trained together with the object arm from the same measured fields.

The approach thus sits between existing ways of controlling light in complex media. A full TM~\cite{popoff2010measuring} is exhaustive but costs at least one pattern per input mode, and it becomes stale when the medium changes unless it is tracked online~\cite{valzania2023online}. Feedback-based shaping~\cite{vellekoop2007focusing} and reinforcement learning~\cite{li2026model} need no model but find the direction of improvement by trial and error. Physics-aware training~\cite{wright2022deep} learns its digital backward pass from large input and output data sets, optical backward passes~\cite{xue2024fully,zhou2026photonic,mididoddi2025threading} use additional optics on the far side of the medium, and physical models fitted to a few measurements have guided wavefront shaping and computational correction~\cite{thendiyammal2020model,matthes2021learning,haim2025image}. Here, one measured field per iteration and a model from a handful of patterns suffice to drive an in situ loop for any loss computed from the camera. Unlike physics-aware training, whose backward pass acts on the field predicted by its twin, the adjoint here acts on the measured complex field, so the model enters the update only through its propagation back to the SLM. On the reference side, interference with a reference has computed optical inner products~\cite{hamerly2019large}, and scattering media have been combined with neural-network decryption~\cite{zhao2022speckle}; we train the reference with a gradient that is independent of the object arm and decrypt in a single read. The method is also simple to adopt, as it needs only the SLM, camera and reference path of an off-axis holographic setup, fits its model in seconds on a standard CPU and costs one SLM pattern per iteration, so it can replace the TM calibration of an existing system without changing its optics.

Each demonstration was designed as a proof of principle and points to a clear next step. The thick diffuser shows that an inexact model still trains the system, forming a letter within 0.6~dB of a dense TM of all 1,024 modes (Supplementary Fig.~S8), whereas refreshing was tested on a thin diffuser, under drift induced by walking the illumination across it. The natural next test is drift in thick media and in multimode fibres, where bending provides a many-parameter drift and the fibre has a compact physical description in its guided modes~\cite{ploschner2015seeing,matthes2021learning}; faster modulators and fitting on dedicated hardware would shorten the refresh, which takes seconds here. Such an experiment would also take the method beyond our current scale of about a thousand input modes, which suits a proof of principle but is not a limit of the method, since the cost of a TM grows with the number of modes, whereas the size of the physical model is set by the bandwidth of the optics, so the advantage should widen as systems grow. In a digital twin of the setup, the ten-probe model keeps a nearly constant held-out field correlation of about 0.85 from 256 to 16,384 SLM cells, whereas a TM needs one to three patterns per cell (Supplementary Note~6, Supplementary Fig.~S9). In the reference arm, the linear diffuser makes the processor a linear classifier, and even the full detected intensity is only a polynomial of degree two in the input, which is not a universal approximator~\cite{leshno1993multilayer}. Universality needs a non-polynomial element, such as detector saturation or a nonlinear medium. Since the reference gradient does not depend on the object arm, a nonlinear multimode fibre could supply a nonlinear projection with no change to the training (Supplementary Note~2A), and the joint training shown here would then train a nonlinear processor end to end. Finally, the encryption was tested only against reads without the keys; until a formal security analysis, including known-plaintext attacks~\cite{peng2006known}, is available, we regard it as a physical two-key scheme rather than a proven cryptosystem.

More broadly, measuring the forward field and modelling only the adjoint applies wherever the output field can be measured and the system has a compact physical description. Multi-plane light converters, mode sorters and diffractive layers, limited in practice by alignment errors, could be trained in situ in the same way. Dynamic media such as biological tissue, atmospheric turbulence and free-space links require the same refresh at higher speed, as does laser beam shaping for material processing, where optics heat up and drift during operation. The two-key scheme, with the diffuser as an unclonable key and the trained reference as a designed one, relates to optical physical unclonable functions~\cite{pappu2002physical,horstmeyer2013physical} and suggests uses in optical authentication.

An off-axis hologram supplies the measured error of every update, a compact physical model learned from a few probe patterns supplies its direction, and because that model can be refreshed within seconds, the system keeps itself calibrated while the medium changes. Training the reference arm turns the same hologram into an optical inner product for computing and encryption. Optical systems need no longer be characterized once and then trusted. They can learn, and recalibrate themselves, while they work.

\section{Methods}\label{sec:methods}

\subsection*{Optical platform}

All experiments were performed on a two-arm off-axis interferometer whose arms are written side by side on a single phase-only SLM, illuminated by a laser at 1,030~nm (Fig.~\ref{fig1}b). The object arm passes through a complex medium before reaching a CMOS camera, whereas the reference arm bypasses the medium and meets the object beam on the camera at a small angle. A 4f relay images both SLM regions, and an iris in its Fourier plane passes only the first diffraction order. Phases were written with a pixelated complex encoding~\cite{arrizon2007pixelated}, which at unit amplitude reduces to a phase on a blazed carrier. The iris therefore doubles as a programmable shutter, since a region written without its blazed carrier sends its light into the blocked zero order. The setup and the encoding are detailed in Supplementary Note~1A.

\subsection*{Measuring the complex field}

Since the reference arrives at an angle, every camera frame is an off-axis hologram, $I=|O+R|^{2}$, whose interference term is separated from the others in the Fourier domain. We isolated this term by Fourier filtering~\cite{cuche2000spatial} and divided it by the measured reference envelope, which yields the complex object field $\hat O$, amplitude and phase, from a single frame. Separating this term from the others with a flat reference requires about 12 camera pixels per input mode~\cite{loterie2017microscopy}. Each hologram records one SLM pattern; how its camera exposures are combined is given in Supplementary Note~1A. A background field that does not depend on the object pattern was measured at the start of each session and removed before any model fitting. Throughout, agreement between two fields is quantified by their gauge-invariant correlation, which ignores the global phase and amplitude of a frame (Supplementary Note~1B).

\subsection*{Training principle and model of the object arm}

Both arms were trained with the gradients of equation~(\ref{eq:gradients}). In each case, the error field is formed from the measured object field and the derivative of the loss with respect to the readout, which is the object intensity for the object arm and the interference term for the reference arm. The two arms differ only in the adjoint that maps this error field back to the SLM. For the reference arm, this adjoint involves free-space optics only, and since the object field does not depend on the reference pattern, a single measurement of each object field suffices to compute the loss and its gradient for any reference. For the object arm, the adjoint passes through the medium and is the only quantity we model; each measured field was aligned with the model by one complex factor per frame before the adjoint was applied. Both gradients match finite differences on random instances to a relative error below $10^{-8}$ (Supplementary Note~2A).

The object arm was described by the compact physical model of equation~(\ref{eq:model}), which comprises free-space propagation through the relay, with the iris and low-order aberrations as a pupil, followed by a complex screen at a fitted distance before the camera, which represents the diffuser (Supplementary Note~3A). Propagation was computed with the band-limited angular-spectrum method~\cite{matsushima2009band}. The model is linear in the SLM field and fully differentiable, so its parameters were fitted to measured fields with exact gradients, that is, by the adjoint method itself, using a quasi-Newton optimizer~\cite{liu1989limited} and a gauge-invariant misfit. The fit is linear in the screen for fixed frame phases and tilts, but not convex as a whole, and was started from the previous screen. Parameters that depend only on the optical configuration were calibrated once, and the diffuser screen was refitted at the start of each session from 10 random probe patterns. Since the illuminating speckle has exponential intensity statistics~\cite{goodman2020speckle}, a region of the screen dark in one probe is lit in another, so a few probes constrain the whole screen.

\subsection*{In situ training of the object arm}

In every iteration, the current pattern was displayed and its field measured, the error field was carried back to the SLM by the model adjoint, and the pattern was updated with Adam~\cite{kingma2014adam}. We applied this loop first to focusing and image formation through the diffuser. For focusing, the loss rewarded the intensity in a disk of one speckle grain at the centre of the field. For image formation, the loss was the energy-normalized mean squared error between the measured intensity and the target, which is insensitive to the overall brightness, and both images, the yin-yang and the clock tower, were trained for 150 iterations (Supplementary Note~3B). The same loop trained the object arm as the input layer of an optical classifier. FashionMNIST images~\cite{xiao2017fashion} were written as phases on the object cells, on top of the trained mask, so that each image passed through the mask and the diffuser before reaching the camera. The predicted class was the brightest of ten detector regions of the measured field, with no digital weights. The mask was trained in situ for eight epochs with the cross-entropy of the detector energies as the loss, and the mask with the best validation accuracy was evaluated once on the test set (Supplementary Note~3B).

\subsection*{Transmission-matrix benchmark}

To compare the few-frame model with an exhaustive one, we measured the TM of the object arm in the same session from $3N$ random patterns, where $N$ is the number of SLM cells of the object pattern~\cite{popoff2010measuring,yu2013measuring}. We used $3N$ random patterns, an overcomplete set that keeps the estimate well conditioned under noise. The TM replaced the physical model in an otherwise identical loop, so that only the adjoint differed. In Fig.~\ref{fig2}, final PSNR values are those of the mean image of ten final frames, whereas the curves of Fig.~\ref{fig2}e give the PSNR read in each iteration. The physical model used 10 probe patterns per session, in addition to 50 patterns used once to calibrate the optical configuration, whereas the TM used 3,072 patterns (Supplementary Note~3C, Supplementary Table~S1).

\subsection*{Offline training of the reference arm}

The model of the reference arm describes its free-space path with a small number of geometric and aberration parameters. Calibrated from a few random reference patterns in seconds, it predicted the measured reference field with a correlation of 0.85 to 0.88, against a repeatability limit of 0.96 to 0.99 (Supplementary Note~4A, Supplementary Fig.~S3a). After changes of the optics, only seven geometric parameters were refitted, with a simplex search~\cite{nelder1965simplex}. For each task, the object fields of all samples were recorded once with a flat reference. The class score was the interference term between the object field and the trained reference, summed over a detector region and read from the raw frame with the fringe-matched masks of equation~(\ref{eq:readout}); this score is an optical inner product in which the reference acts as a complex weight layer. A raw frame holds four terms, $|O|^{2}+|R|^{2}+OR^{*}+O^{*}R$. Since the reference meets the object beam at an angle, $OR^{*}$ oscillates across the camera at the carrier frequency $-\mathbf{f}_0$, $O^{*}R$ at $+\mathbf{f}_0$, and the two intensities stay near zero frequency, within a band narrower than $|\mathbf{f}_0|$. Each mask is a square detector window $M_c$ multiplied by the complex fringe $e^{j2\pi\mathbf{f}_0\cdot\mathbf{x}}$, which shifts $OR^{*}$ to zero frequency and the other terms to $\mathbf{f}_0$ and $2\mathbf{f}_0$. Summed over a window of about 17 fringe periods, only $OR^{*}$ survives, and the phase $e^{-j\phi}$ between the arms selects its real part. The masks are fixed, so a score costs one multiply and sum over the raw frame, equivalent to a detector matched to the fringes, and needs neither a Fourier transform nor subtraction of the reference intensity. On the bench, raw-frame and demodulated scores of the same frames correlated with $|r|\ge0.999$ (Supplementary Note~4B, Supplementary Fig.~S3b). Alternatively, the scores can be computed from the demodulated field, at the cost of one Fourier filtering and one reference-only frame per read. Fig.~\ref{fig3}c shows this demodulated read for Basic, and Fig.~\ref{fig3}d the all-optical read for the harder Defend the Center; both reads were computed for both scenarios (Supplementary Note~4B). The reference was trained in software on the recorded fields, displayed once and tested with one read per test sample. A data set therefore costs one SLM pattern per training sample and one per test sample, however many training steps are taken digitally. In game play, frames of the ViZDoom~\cite{kempka2016vizdoom} scenarios Basic and Defend the Center were encoded as phase patterns by a fixed colour projection and pooling, without learned features, and an oracle labelled each frame with one of three actions. After training, the system played the game in closed loop on episodes not seen during training (Supplementary Note~4B).

\subsection*{Two-key encryption}

A QR code was stored in the sign of the interference term in each of its modules. The designed key, a reference pattern, was obtained offline in the reference arm, and the ciphertext, a phase pattern on the object cells, was optimized through the diffuser in the object arm, first through the ten-probe model and then in situ; each code was carried by one ciphertext frame. Decryption requires one read with both keys, the diffuser and the reference pattern. Reads with a flat reference, a wrong key or the intensity alone served as attacks (Supplementary Note~4C).

\subsection*{Joint training}

On Defend the Center, both arms were also trained together. Every epoch alternated an in situ update of the object pattern through the ten-probe model, refitted on the fields of each epoch, an offline update of the reference on the fields just measured, and an all-optical validation read from the raw frame; the epoch with the best balanced validation accuracy was kept and tested once (Supplementary Note~4D).

\subsection*{Model refresh under a drifting diffuser}

To test whether the few-frame model can follow a medium that changes during the optimization, the diffuser was made to drift while the object arm was trained. A phase ramp, hidden from the model, moved the illumination across the diffuser and progressively reduced the correlation of the illuminated region with its initial state. The output tilt and shift caused by the ramp were calibrated at the start of the session and removed from the measured fields, so that only the change of the medium seen by the beam remained. The learning rate of the training loop decayed to 0.1 in the refreshed run and to 0.03 in the frozen and drift-free runs. The model was either kept frozen or refreshed at regular intervals from a few random probe patterns and the latest training frame (Supplementary Note~5A).

\subsection*{Training through a thick diffuser}

To test training with an inexact adjoint, the diffuser was replaced by a thick diffuser made of two parafilm-coated glass slides about 15~mm apart. Its memory effect was comparable to the angular spacing of the input modes, so it cannot be described by a single thin screen~\cite{freund1988memory,osnabrugge2017generalized}. We modelled it by two thin screens fitted from 200 probe patterns and repeated the focusing and image-formation experiments, with a grating and EMNIST letters~\cite{cohen2017emnist} as targets, with this deliberately inexact adjoint (Supplementary Note~5B).

\subsection*{Metrics and statistics}

Model accuracy, image quality, enhancement, classification and encryption metrics, and the error bars reported in the figures, are defined in Supplementary Notes~1B and 7. Unless stated otherwise, each condition was measured once in the session shown.

\backmatter

\bmhead{Supplementary information}

Supplementary Notes 1--7, Supplementary Figs.~S1--S9 and Supplementary Table~S1.

\bmhead{Funding}
This work is supported by the Scientific and Technological Research Council of T\"{u}rkiye (T\"{U}B\.{I}TAK) under grant number 122C150.
\bmhead{Competing interests}

The authors declare no competing interests.

\bmhead{Data and code availability}

Data underlying the figures, together with the digital twin results, are publicly available at Zenodo~\cite{ekec2026supplementary}. The scripts that generate them are publicly available on GitHub~\cite{ekec2026holographicgradients}. The datasets used in this study are publicly available from their original sources, namely FashionMNIST~\cite{xiao2017fashion}, EMNIST~\cite{cohen2017emnist} and the ViZDoom platform~\cite{kempka2016vizdoom}. Any additional data generated during the experiments are available from the corresponding author upon reasonable request.

\bmhead{Author contributions}

K.E. and B.U.K. built the digital twin and the experimental setup, and conducted the experiments. U.T. supervised the project. All authors participated in the analysis of the results and the writing of the manuscript.

\bibliography{references}

\begin{thebibliography}{10}
\expandafter\ifx\csname url\endcsname\relax
  \def\url#1{\burl{#1}}\fi
\expandafter\ifx\csname urlprefix\endcsname\relax\def\urlprefix{URL }\fi
\providecommand{\bibinfo}[2]{#2}
\providecommand{\eprint}[2][]{\url{#2}}
\providecommand{\doi}[1]{\url{https://doi.org/#1}}
\bibcommenthead

\bibitem{mosk2012controlling}
\bibinfo{author}{Mosk, A.~P.}, \bibinfo{author}{Lagendijk, A.},
  \bibinfo{author}{Lerosey, G.} \& \bibinfo{author}{Fink, M.}
\newblock \bibinfo{title}{Controlling waves in space and time for imaging and
  focusing in complex media}.
\newblock \emph{\bibinfo{journal}{Nature Photonics}}
  \textbf{\bibinfo{volume}{6}}, \bibinfo{pages}{283--292}
  (\bibinfo{year}{2012}).

\bibitem{rotter2017light}
\bibinfo{author}{Rotter, S.} \& \bibinfo{author}{Gigan, S.}
\newblock \bibinfo{title}{Light fields in complex media: {Mesoscopic}
  scattering meets wave control}.
\newblock \emph{\bibinfo{journal}{Reviews of Modern Physics}}
  \textbf{\bibinfo{volume}{89}}, \bibinfo{pages}{015005}
  (\bibinfo{year}{2017}).

\bibitem{vellekoop2007focusing}
\bibinfo{author}{Vellekoop, I.~M.} \& \bibinfo{author}{Mosk, A.~P.}
\newblock \bibinfo{title}{Focusing coherent light through opaque strongly
  scattering media}.
\newblock \emph{\bibinfo{journal}{Optics Letters}}
  \textbf{\bibinfo{volume}{32}}, \bibinfo{pages}{2309} (\bibinfo{year}{2007}).

\bibitem{popoff2010measuring}
\bibinfo{author}{Popoff, S.~M.} \emph{et~al.}
\newblock \bibinfo{title}{Measuring the transmission matrix in optics: An
  approach to the study and control of light propagation in disordered media}.
\newblock \emph{\bibinfo{journal}{Physical Review Letters}}
  \textbf{\bibinfo{volume}{104}}, \bibinfo{pages}{100601}
  (\bibinfo{year}{2010}).

\bibitem{yu2013measuring}
\bibinfo{author}{Yu, H.} \emph{et~al.}
\newblock \bibinfo{title}{Measuring large optical transmission matrices of
  disordered media}.
\newblock \emph{\bibinfo{journal}{Physical Review Letters}}
  \textbf{\bibinfo{volume}{111}}, \bibinfo{pages}{153902}
  (\bibinfo{year}{2013}).

\bibitem{yaqoob2008optical}
\bibinfo{author}{Yaqoob, Z.}, \bibinfo{author}{Psaltis, D.},
  \bibinfo{author}{Feld, M.~S.} \& \bibinfo{author}{Yang, C.}
\newblock \bibinfo{title}{Optical phase conjugation for turbidity suppression
  in biological samples}.
\newblock \emph{\bibinfo{journal}{Nature Photonics}}
  \textbf{\bibinfo{volume}{2}}, \bibinfo{pages}{110--115}
  (\bibinfo{year}{2008}).

\bibitem{wetzstein2020inference}
\bibinfo{author}{Wetzstein, G.} \emph{et~al.}
\newblock \bibinfo{title}{Inference in artificial intelligence with deep optics
  and photonics}.
\newblock \emph{\bibinfo{journal}{Nature}} \textbf{\bibinfo{volume}{588}},
  \bibinfo{pages}{39--47} (\bibinfo{year}{2020}).

\bibitem{mcmahon2023physics}
\bibinfo{author}{McMahon, P.~L.}
\newblock \bibinfo{title}{The physics of optical computing}.
\newblock \emph{\bibinfo{journal}{Nature Reviews Physics}}
  \textbf{\bibinfo{volume}{5}}, \bibinfo{pages}{717--734}
  (\bibinfo{year}{2023}).

\bibitem{lin2018all}
\bibinfo{author}{Lin, X.} \emph{et~al.}
\newblock \bibinfo{title}{All-optical machine learning using diffractive deep
  neural networks}.
\newblock \emph{\bibinfo{journal}{Science}} \textbf{\bibinfo{volume}{361}},
  \bibinfo{pages}{1004--1008} (\bibinfo{year}{2018}).

\bibitem{hamerly2019large}
\bibinfo{author}{Hamerly, R.}, \bibinfo{author}{Bernstein, L.},
  \bibinfo{author}{Sludds, A.}, \bibinfo{author}{Solja{\v{c}}i{\'c}, M.} \&
  \bibinfo{author}{Englund, D.}
\newblock \bibinfo{title}{{Large-Scale} optical neural networks based on
  photoelectric multiplication}.
\newblock \emph{\bibinfo{journal}{Physical Review X}}
  \textbf{\bibinfo{volume}{9}}, \bibinfo{pages}{021032} (\bibinfo{year}{2019}).

\bibitem{shen2017deep}
\bibinfo{author}{Shen, Y.} \emph{et~al.}
\newblock \bibinfo{title}{Deep learning with coherent nanophotonic circuits}.
\newblock \emph{\bibinfo{journal}{Nature Photonics}}
  \textbf{\bibinfo{volume}{11}}, \bibinfo{pages}{441--446}
  (\bibinfo{year}{2017}).

\bibitem{carpinlioglu2025genetically}
\bibinfo{author}{{\c{C}}arp{\i}nl{\i}o{\u{g}}lu, B.} \&
  \bibinfo{author}{Te{\u{g}}in, U.}
\newblock \bibinfo{title}{Genetically programmable optical random neural
  networks}.
\newblock \emph{\bibinfo{journal}{Communications Physics}}
  \textbf{\bibinfo{volume}{8}}, \bibinfo{pages}{349} (\bibinfo{year}{2025}).

\bibitem{rafayelyan2020large}
\bibinfo{author}{Rafayelyan, M.}, \bibinfo{author}{Dong, J.},
  \bibinfo{author}{Tan, Y.}, \bibinfo{author}{Krzakala, F.} \&
  \bibinfo{author}{Gigan, S.}
\newblock \bibinfo{title}{{Large-Scale} optical reservoir computing for
  spatiotemporal chaotic systems prediction}.
\newblock \emph{\bibinfo{journal}{Physical Review X}}
  \textbf{\bibinfo{volume}{10}}, \bibinfo{pages}{041037}
  (\bibinfo{year}{2020}).

\bibitem{saade2016random}
\bibinfo{author}{Saade, A.} \emph{et~al.}
\newblock \bibinfo{title}{Random projections through multiple optical
  scattering: {Approximating} {Kernels} at the speed of light}.
\newblock \emph{\bibinfo{journal}{Proc. 2016 IEEE International Conference on
  Acoustics, Speech and Signal Processing (ICASSP)}}
  \bibinfo{pages}{6215--6219} (\bibinfo{year}{2016}).

\bibitem{tegin2021scalable}
\bibinfo{author}{Te{\u{g}}in, U.}, \bibinfo{author}{Y{\i}ld{\i}r{\i}m, M.},
  \bibinfo{author}{O{\u{g}}uz, {\.I}.}, \bibinfo{author}{Moser, C.} \&
  \bibinfo{author}{Psaltis, D.}
\newblock \bibinfo{title}{Scalable optical learning operator}.
\newblock \emph{\bibinfo{journal}{Nature Computational Science}}
  \textbf{\bibinfo{volume}{1}}, \bibinfo{pages}{542--549}
  (\bibinfo{year}{2021}).

\bibitem{oguz2025training}
\bibinfo{author}{Oguz, I.} \emph{et~al.}
\newblock \bibinfo{title}{Training hybrid neural networks with multimode
  optical nonlinearities using digital twins}.
\newblock \emph{\bibinfo{journal}{Nanophotonics}}
  \textbf{\bibinfo{volume}{14}}, \bibinfo{pages}{2787--2797}
  (\bibinfo{year}{2025}).

\bibitem{kesgin2025photonic}
\bibinfo{author}{Kesgin, B.~U.} \& \bibinfo{author}{Te{\u{g}}in, U.}
\newblock \bibinfo{title}{Photonic neural networks at the edge of
  spatiotemporal chaos in multimode fibers}.
\newblock \emph{\bibinfo{journal}{Nanophotonics}}
  \textbf{\bibinfo{volume}{14}}, \bibinfo{pages}{2723--2732}
  (\bibinfo{year}{2025}).

\bibitem{eslik2026recurrent}
\bibinfo{author}{Eslik, D.}, \bibinfo{author}{Kesgin, B.~U.} \&
  \bibinfo{author}{Te{\u{g}}in, U.}
\newblock \bibinfo{title}{Recurrent neural networks implemented through
  spatiotemporal light propagation in optical fibers}.
\newblock \emph{\bibinfo{journal}{Optica}}  (\bibinfo{year}{2026}).

\bibitem{wright2022deep}
\bibinfo{author}{Wright, L.~G.} \emph{et~al.}
\newblock \bibinfo{title}{Deep physical neural networks trained with
  backpropagation}.
\newblock \emph{\bibinfo{journal}{Nature}} \textbf{\bibinfo{volume}{601}},
  \bibinfo{pages}{549--555} (\bibinfo{year}{2022}).

\bibitem{momeni2025training}
\bibinfo{author}{Momeni, A.} \emph{et~al.}
\newblock \bibinfo{title}{Training of physical neural networks}.
\newblock \emph{\bibinfo{journal}{Nature}} \textbf{\bibinfo{volume}{645}},
  \bibinfo{pages}{53--61} (\bibinfo{year}{2025}).

\bibitem{li2018deep}
\bibinfo{author}{Li, Y.}, \bibinfo{author}{Xue, Y.} \& \bibinfo{author}{Tian,
  L.}
\newblock \bibinfo{title}{Deep speckle correlation: a deep learning approach
  toward scalable imaging through scattering media}.
\newblock \emph{\bibinfo{journal}{Optica}} \textbf{\bibinfo{volume}{5}},
  \bibinfo{pages}{1181} (\bibinfo{year}{2018}).

\bibitem{ekec2026resolution}
\bibinfo{author}{Ekec, K.} \& \bibinfo{author}{Te{\u{g}}in, U.}
\newblock \bibinfo{title}{Resolution-agnostic lensless imaging via {Fourier}
  neural operators}.
\newblock \emph{\bibinfo{journal}{Optics Express}}
  \textbf{\bibinfo{volume}{34}}, \bibinfo{pages}{34757--34772}
  (\bibinfo{year}{2026}).

\bibitem{li2026model}
\bibinfo{author}{Li, Y.}, \bibinfo{author}{Chen, S.}, \bibinfo{author}{Gong,
  T.} \& \bibinfo{author}{Ozcan, A.}
\newblock \bibinfo{title}{Model-free optical processors using in situ
  reinforcement learning with proximal policy optimization}.
\newblock \emph{\bibinfo{journal}{Light: Science \& Applications}}
  \textbf{\bibinfo{volume}{15}}, \bibinfo{pages}{32} (\bibinfo{year}{2026}).

\bibitem{liu2026focusing}
\bibinfo{author}{Liu, Y.} \emph{et~al.}
\newblock \bibinfo{title}{Focusing light through turbid and thick scattering
  media via a hybrid genetic neural network algorithm}.
\newblock \emph{\bibinfo{journal}{Photonics Research}}
  \textbf{\bibinfo{volume}{14}}, \bibinfo{pages}{2306} (\bibinfo{year}{2026}).

\bibitem{kesgin2025fiber}
\bibinfo{author}{Kesgin, B.~U.}, \bibinfo{author}{Y{\"u}ce, F.} \&
  \bibinfo{author}{Te{\u{g}}in, U.}
\newblock \bibinfo{title}{Fiber-based diffractive deep neural network}.
\newblock \emph{\bibinfo{journal}{Optics Letters}}
  \textbf{\bibinfo{volume}{50}}, \bibinfo{pages}{5254} (\bibinfo{year}{2025}).

\bibitem{momeni2023backpropagation}
\bibinfo{author}{Momeni, A.}, \bibinfo{author}{Rahmani, B.},
  \bibinfo{author}{Mall{\'e}jac, M.}, \bibinfo{author}{del Hougne, P.} \&
  \bibinfo{author}{Fleury, R.}
\newblock \bibinfo{title}{Backpropagation-free training of deep physical neural
  networks}.
\newblock \emph{\bibinfo{journal}{Science}} \textbf{\bibinfo{volume}{382}},
  \bibinfo{pages}{1297--1303} (\bibinfo{year}{2023}).

\bibitem{oguz2023forward}
\bibinfo{author}{Oguz, I.} \emph{et~al.}
\newblock \bibinfo{title}{Forward--forward training of an optical neural
  network}.
\newblock \emph{\bibinfo{journal}{Optics Letters}}
  \textbf{\bibinfo{volume}{48}}, \bibinfo{pages}{5249} (\bibinfo{year}{2023}).

\bibitem{kesgin2026optical}
\bibinfo{author}{Kesgin, B.~U.}, \bibinfo{author}{Durdu, G.~Y.} \&
  \bibinfo{author}{Te{\u{g}}in, U.}
\newblock \bibinfo{title}{Optical spiking neural networks via rogue-wave
  statistics}.
\newblock \emph{\bibinfo{journal}{npj Unconventional Computing}}
  \textbf{\bibinfo{volume}{3}}, \bibinfo{pages}{33} (\bibinfo{year}{2026}).

\bibitem{wang2026tron}
\bibinfo{author}{Wang, Z.} \emph{et~al.}
\newblock \bibinfo{title}{{TRON}: {Trainable}, architecture-reconfigurable
  random optical neural networks}.
\newblock \bibinfo{howpublished}{arXiv preprint arXiv:2604.16228}
  (\bibinfo{year}{2026}).
\newblock
  \bibinfo{eprint}{{\href{https://arxiv.org/abs/2604.16228}{{arXiv:2604.16228}}}}.

\bibitem{li2018imaging}
\bibinfo{author}{Li, S.}, \bibinfo{author}{Deng, M.}, \bibinfo{author}{Lee,
  J.}, \bibinfo{author}{Sinha, A.} \& \bibinfo{author}{Barbastathis, G.}
\newblock \bibinfo{title}{Imaging through glass diffusers using densely
  connected convolutional networks}.
\newblock \emph{\bibinfo{journal}{Optica}} \textbf{\bibinfo{volume}{5}},
  \bibinfo{pages}{803} (\bibinfo{year}{2018}).

\bibitem{matthes2021learning}
\bibinfo{author}{Matth{\`e}s, M.~W.}, \bibinfo{author}{Bromberg, Y.},
  \bibinfo{author}{de~Rosny, J.} \& \bibinfo{author}{Popoff, S.~M.}
\newblock \bibinfo{title}{Learning and avoiding disorder in multimode fibers}.
\newblock \emph{\bibinfo{journal}{Physical Review X}}
  \textbf{\bibinfo{volume}{11}}, \bibinfo{pages}{021060}
  (\bibinfo{year}{2021}).

\bibitem{hughes2018training}
\bibinfo{author}{Hughes, T.~W.}, \bibinfo{author}{Minkov, M.},
  \bibinfo{author}{Shi, Y.} \& \bibinfo{author}{Fan, S.}
\newblock \bibinfo{title}{Training of photonic neural networks through in situ
  backpropagation and gradient measurement}.
\newblock \emph{\bibinfo{journal}{Optica}} \textbf{\bibinfo{volume}{5}},
  \bibinfo{pages}{864} (\bibinfo{year}{2018}).

\bibitem{pai2023experimentally}
\bibinfo{author}{Pai, S.} \emph{et~al.}
\newblock \bibinfo{title}{Experimentally realized in situ backpropagation for
  deep learning in photonic neural networks}.
\newblock \emph{\bibinfo{journal}{Science}} \textbf{\bibinfo{volume}{380}},
  \bibinfo{pages}{398--404} (\bibinfo{year}{2023}).

\bibitem{spall2025training}
\bibinfo{author}{Spall, J.}, \bibinfo{author}{Guo, X.} \&
  \bibinfo{author}{Lvovsky, A.~I.}
\newblock \bibinfo{title}{Training neural networks with end-to-end optical
  backpropagation}.
\newblock \emph{\bibinfo{journal}{Advanced Photonics}}
  \textbf{\bibinfo{volume}{7}}, \bibinfo{pages}{016004} (\bibinfo{year}{2025}).

\bibitem{monin2026rapid}
\bibinfo{author}{Monin, S.}, \bibinfo{author}{Alterman, M.} \&
  \bibinfo{author}{Levin, A.}
\newblock \bibinfo{title}{Rapid wavefront shaping using an optical gradient
  acquisition}.
\newblock \emph{\bibinfo{journal}{Nature Communications}}
  \textbf{\bibinfo{volume}{17}}, \bibinfo{pages}{1537} (\bibinfo{year}{2026}).

\bibitem{freund1988memory}
\bibinfo{author}{Freund, I.}, \bibinfo{author}{Rosenbluh, M.} \&
  \bibinfo{author}{Feng, S.}
\newblock \bibinfo{title}{Memory effects in propagation of optical waves
  through disordered media}.
\newblock \emph{\bibinfo{journal}{Physical Review Letters}}
  \textbf{\bibinfo{volume}{61}}, \bibinfo{pages}{2328--2331}
  (\bibinfo{year}{1988}).

\bibitem{bertolotti2012non}
\bibinfo{author}{Bertolotti, J.} \emph{et~al.}
\newblock \bibinfo{title}{Non-invasive imaging through opaque scattering
  layers}.
\newblock \emph{\bibinfo{journal}{Nature}} \textbf{\bibinfo{volume}{491}},
  \bibinfo{pages}{232--234} (\bibinfo{year}{2012}).

\bibitem{thendiyammal2020model}
\bibinfo{author}{Thendiyammal, A.}, \bibinfo{author}{Osnabrugge, G.},
  \bibinfo{author}{Knop, T.} \& \bibinfo{author}{Vellekoop, I.~M.}
\newblock \bibinfo{title}{Model-based wavefront shaping microscopy}.
\newblock \emph{\bibinfo{journal}{Optics Letters}}
  \textbf{\bibinfo{volume}{45}}, \bibinfo{pages}{5101} (\bibinfo{year}{2020}).

\bibitem{haim2025image}
\bibinfo{author}{Haim, O.}, \bibinfo{author}{Boger-Lombard, J.} \&
  \bibinfo{author}{Katz, O.}
\newblock \bibinfo{title}{Image-guided computational holographic wavefront
  shaping}.
\newblock \emph{\bibinfo{journal}{Nature Photonics}}
  \textbf{\bibinfo{volume}{19}}, \bibinfo{pages}{44--53}
  (\bibinfo{year}{2025}).

\bibitem{fienup1982phase}
\bibinfo{author}{Fienup, J.~R.}
\newblock \bibinfo{title}{Phase retrieval algorithms: a comparison}.
\newblock \emph{\bibinfo{journal}{Applied Optics}}
  \textbf{\bibinfo{volume}{21}}, \bibinfo{pages}{2758} (\bibinfo{year}{1982}).

\bibitem{dremeau2015reference}
\bibinfo{author}{Dr{\'e}meau, A.} \emph{et~al.}
\newblock \bibinfo{title}{Reference-less measurement of the transmission matrix
  of a highly scattering material using a {DMD} and phase retrieval
  techniques}.
\newblock \emph{\bibinfo{journal}{Optics Express}}
  \textbf{\bibinfo{volume}{23}}, \bibinfo{pages}{11898} (\bibinfo{year}{2015}).

\bibitem{xue2024fully}
\bibinfo{author}{Xue, Z.} \emph{et~al.}
\newblock \bibinfo{title}{Fully forward mode training for optical neural
  networks}.
\newblock \emph{\bibinfo{journal}{Nature}} \textbf{\bibinfo{volume}{632}},
  \bibinfo{pages}{280--286} (\bibinfo{year}{2024}).

\bibitem{zhou2026photonic}
\bibinfo{author}{Zhou, T.} \emph{et~al.}
\newblock \bibinfo{title}{Photonic neuromorphic learning via generalized in
  situ physical gradient descent}.
\newblock \emph{\bibinfo{journal}{Nature Computational Science}}
  (\bibinfo{year}{2026}).

\bibitem{mididoddi2025threading}
\bibinfo{author}{Mididoddi, C.~K.} \emph{et~al.}
\newblock \bibinfo{title}{Threading light through dynamic complex media}.
\newblock \emph{\bibinfo{journal}{Nature Photonics}}
  \textbf{\bibinfo{volume}{19}}, \bibinfo{pages}{434--440}
  (\bibinfo{year}{2025}).

\bibitem{leith1962reconstructed}
\bibinfo{author}{Leith, E.~N.} \& \bibinfo{author}{Upatnieks, J.}
\newblock \bibinfo{title}{Reconstructed wavefronts and communication theory}.
\newblock \emph{\bibinfo{journal}{Journal of the Optical Society of America}}
  \textbf{\bibinfo{volume}{52}}, \bibinfo{pages}{1123} (\bibinfo{year}{1962}).

\bibitem{cuche2000spatial}
\bibinfo{author}{Cuche, E.}, \bibinfo{author}{Marquet, P.} \&
  \bibinfo{author}{Depeursinge, C.}
\newblock \bibinfo{title}{Spatial filtering for zero-order and twin-image
  elimination in digital off-axis holography}.
\newblock \emph{\bibinfo{journal}{Applied Optics}}
  \textbf{\bibinfo{volume}{39}}, \bibinfo{pages}{4070} (\bibinfo{year}{2000}).

\bibitem{chen2023deep}
\bibinfo{author}{Chen, Z.} \emph{et~al.}
\newblock \bibinfo{title}{Deep learning with coherent {VCSEL} neural networks}.
\newblock \emph{\bibinfo{journal}{Nature Photonics}}
  \textbf{\bibinfo{volume}{17}}, \bibinfo{pages}{723--730}
  (\bibinfo{year}{2023}).

\bibitem{wirtinger1927formalen}
\bibinfo{author}{Wirtinger, W.}
\newblock \bibinfo{title}{Zur formalen {Theorie} der {Funktionen} von mehr
  komplexen {Ver{\"a}nderlichen}}.
\newblock \emph{\bibinfo{journal}{Mathematische Annalen}}
  \textbf{\bibinfo{volume}{97}}, \bibinfo{pages}{357--375}
  (\bibinfo{year}{1927}).

\bibitem{haskel2018modeling}
\bibinfo{author}{Haskel, M.} \& \bibinfo{author}{Stern, A.}
\newblock \bibinfo{title}{Modeling optical memory effects with phase screens}.
\newblock \emph{\bibinfo{journal}{Optics Express}}
  \textbf{\bibinfo{volume}{26}}, \bibinfo{pages}{29231} (\bibinfo{year}{2018}).

\bibitem{xiao2017fashion}
\bibinfo{author}{Xiao, H.}, \bibinfo{author}{Rasul, K.} \&
  \bibinfo{author}{Vollgraf, R.}
\newblock \bibinfo{title}{{Fashion-MNIST}: a novel image dataset for
  benchmarking machine learning algorithms}.
\newblock \bibinfo{howpublished}{arXiv preprint arXiv:1708.07747}
  (\bibinfo{year}{2017}).
\newblock
  \bibinfo{eprint}{{\href{https://arxiv.org/abs/1708.07747}{{arXiv:1708.07747}}}}.

\bibitem{kempka2016vizdoom}
\bibinfo{author}{Kempka, M.}, \bibinfo{author}{Wydmuch, M.},
  \bibinfo{author}{Runc, G.}, \bibinfo{author}{Toczek, J.} \&
  \bibinfo{author}{Ja{\'s}kowski, W.}
\newblock \bibinfo{title}{{ViZDoom}: {A} {Doom-based} {AI} research platform
  for visual reinforcement learning}.
\newblock \emph{\bibinfo{journal}{Proc. 2016 IEEE Conference on Computational
  Intelligence and Games (CIG)}} \bibinfo{pages}{1--8} (\bibinfo{year}{2016}).

\bibitem{refregier1995optical}
\bibinfo{author}{Refregier, P.} \& \bibinfo{author}{Javidi, B.}
\newblock \bibinfo{title}{Optical image encryption based on input plane and
  {Fourier} plane random encoding}.
\newblock \emph{\bibinfo{journal}{Optics Letters}}
  \textbf{\bibinfo{volume}{20}}, \bibinfo{pages}{767} (\bibinfo{year}{1995}).

\bibitem{horstmeyer2013physical}
\bibinfo{author}{Horstmeyer, R.}, \bibinfo{author}{Judkewitz, B.},
  \bibinfo{author}{Vellekoop, I.~M.}, \bibinfo{author}{Assawaworrarit, S.} \&
  \bibinfo{author}{Yang, C.}
\newblock \bibinfo{title}{Physical key-protected one-time pad}.
\newblock \emph{\bibinfo{journal}{Scientific Reports}}
  \textbf{\bibinfo{volume}{3}}, \bibinfo{pages}{3543} (\bibinfo{year}{2013}).

\bibitem{barrera2013optical}
\bibinfo{author}{Barrera, J.~F.}, \bibinfo{author}{Mira, A.} \&
  \bibinfo{author}{Torroba, R.}
\newblock \bibinfo{title}{Optical encryption and {QR} codes: {Secure} and
  noise-free information retrieval}.
\newblock \emph{\bibinfo{journal}{Optics Express}}
  \textbf{\bibinfo{volume}{21}}, \bibinfo{pages}{5373} (\bibinfo{year}{2013}).

\bibitem{osnabrugge2017generalized}
\bibinfo{author}{Osnabrugge, G.}, \bibinfo{author}{Horstmeyer, R.},
  \bibinfo{author}{Papadopoulos, I.~N.}, \bibinfo{author}{Judkewitz, B.} \&
  \bibinfo{author}{Vellekoop, I.~M.}
\newblock \bibinfo{title}{Generalized optical memory effect}.
\newblock \emph{\bibinfo{journal}{Optica}} \textbf{\bibinfo{volume}{4}},
  \bibinfo{pages}{886} (\bibinfo{year}{2017}).

\bibitem{kamilov2015learning}
\bibinfo{author}{Kamilov, U.~S.} \emph{et~al.}
\newblock \bibinfo{title}{Learning approach to optical tomography}.
\newblock \emph{\bibinfo{journal}{Optica}} \textbf{\bibinfo{volume}{2}},
  \bibinfo{pages}{517} (\bibinfo{year}{2015}).

\bibitem{lib2020real}
\bibinfo{author}{Lib, O.}, \bibinfo{author}{Hasson, G.} \&
  \bibinfo{author}{Bromberg, Y.}
\newblock \bibinfo{title}{Real-time shaping of entangled photons by classical
  control and feedback}.
\newblock \emph{\bibinfo{journal}{Science Advances}}
  \textbf{\bibinfo{volume}{6}}, \bibinfo{pages}{eabb6298}
  (\bibinfo{year}{2020}).

\bibitem{valzania2023online}
\bibinfo{author}{Valzania, L.} \& \bibinfo{author}{Gigan, S.}
\newblock \bibinfo{title}{Online learning of the transmission matrix of dynamic
  scattering media}.
\newblock \emph{\bibinfo{journal}{Optica}} \textbf{\bibinfo{volume}{10}},
  \bibinfo{pages}{708} (\bibinfo{year}{2023}).

\bibitem{zhao2022speckle}
\bibinfo{author}{Zhao, Q.} \emph{et~al.}
\newblock \bibinfo{title}{{Speckle-Based} optical cryptosystem and its
  application for human face recognition via deep learning}.
\newblock \emph{\bibinfo{journal}{Advanced Science}}
  \textbf{\bibinfo{volume}{9}}, \bibinfo{pages}{2202407}
  (\bibinfo{year}{2022}).

\bibitem{ploschner2015seeing}
\bibinfo{author}{Pl{\"o}schner, M.}, \bibinfo{author}{Tyc, T.} \&
  \bibinfo{author}{{\v{C}}i{\v{z}}m{\'a}r, T.}
\newblock \bibinfo{title}{Seeing through chaos in multimode fibres}.
\newblock \emph{\bibinfo{journal}{Nature Photonics}}
  \textbf{\bibinfo{volume}{9}}, \bibinfo{pages}{529--535}
  (\bibinfo{year}{2015}).

\bibitem{leshno1993multilayer}
\bibinfo{author}{Leshno, M.}, \bibinfo{author}{Lin, V.~Y.},
  \bibinfo{author}{Pinkus, A.} \& \bibinfo{author}{Schocken, S.}
\newblock \bibinfo{title}{Multilayer feedforward networks with a nonpolynomial
  activation function can approximate any function}.
\newblock \emph{\bibinfo{journal}{Neural Networks}}
  \textbf{\bibinfo{volume}{6}}, \bibinfo{pages}{861--867}
  (\bibinfo{year}{1993}).

\bibitem{peng2006known}
\bibinfo{author}{Peng, X.}, \bibinfo{author}{Zhang, P.}, \bibinfo{author}{Wei,
  H.} \& \bibinfo{author}{Yu, B.}
\newblock \bibinfo{title}{Known-plaintext attack on optical encryption based on
  double random phase keys}.
\newblock \emph{\bibinfo{journal}{Optics Letters}}
  \textbf{\bibinfo{volume}{31}}, \bibinfo{pages}{1044} (\bibinfo{year}{2006}).

\bibitem{pappu2002physical}
\bibinfo{author}{Pappu, R.}, \bibinfo{author}{Recht, B.},
  \bibinfo{author}{Taylor, J.} \& \bibinfo{author}{Gershenfeld, N.}
\newblock \bibinfo{title}{Physical {One-Way} functions}.
\newblock \emph{\bibinfo{journal}{Science}} \textbf{\bibinfo{volume}{297}},
  \bibinfo{pages}{2026--2030} (\bibinfo{year}{2002}).

\bibitem{arrizon2007pixelated}
\bibinfo{author}{Arriz{\'o}n, V.}, \bibinfo{author}{Ruiz, U.},
  \bibinfo{author}{Carrada, R.} \& \bibinfo{author}{Gonz{\'a}lez, L.~A.}
\newblock \bibinfo{title}{Pixelated phase computer holograms for the accurate
  encoding of scalar complex fields}.
\newblock \emph{\bibinfo{journal}{Journal of the Optical Society of America A}}
  \textbf{\bibinfo{volume}{24}}, \bibinfo{pages}{3500--3507}
  (\bibinfo{year}{2007}).

\bibitem{loterie2017microscopy}
\bibinfo{author}{Loterie, D. C.-M.}
\newblock \emph{\bibinfo{title}{Microscopy and digital light shaping through
  optical fibers}}.
\newblock Ph.D. thesis, \bibinfo{school}{EPFL}, \bibinfo{address}{Lausanne}
  (\bibinfo{year}{2017}).
\newblock \bibinfo{note}{EPFL thesis no. 7799}.

\bibitem{matsushima2009band}
\bibinfo{author}{Matsushima, K.} \& \bibinfo{author}{Shimobaba, T.}
\newblock \bibinfo{title}{Band-limited angular spectrum method for numerical
  simulation of free-space propagation in far and near fields}.
\newblock \emph{\bibinfo{journal}{Optics Express}}
  \textbf{\bibinfo{volume}{17}}, \bibinfo{pages}{19662} (\bibinfo{year}{2009}).

\bibitem{liu1989limited}
\bibinfo{author}{Liu, D.~C.} \& \bibinfo{author}{Nocedal, J.}
\newblock \bibinfo{title}{On the limited memory {BFGS} method for large scale
  optimization}.
\newblock \emph{\bibinfo{journal}{Mathematical Programming}}
  \textbf{\bibinfo{volume}{45}}, \bibinfo{pages}{503--528}
  (\bibinfo{year}{1989}).

\bibitem{goodman2020speckle}
\bibinfo{author}{Goodman, J.~W.}
\newblock \emph{\bibinfo{title}{Speckle Phenomena in Optics: Theory and
  Applications, Second Edition}}  (\bibinfo{publisher}{SPIE},
  \bibinfo{year}{2020}).

\bibitem{kingma2014adam}
\bibinfo{author}{Kingma, D.~P.} \& \bibinfo{author}{Ba, J.}
\newblock \bibinfo{title}{Adam: A method for stochastic optimization}.
\newblock \emph{\bibinfo{journal}{Proc. 3rd International Conference on
  Learning Representations (ICLR)}}  (\bibinfo{year}{2015}).

\bibitem{nelder1965simplex}
\bibinfo{author}{Nelder, J.~A.} \& \bibinfo{author}{Mead, R.}
\newblock \bibinfo{title}{A simplex method for function minimization}.
\newblock \emph{\bibinfo{journal}{The Computer Journal}}
  \textbf{\bibinfo{volume}{7}}, \bibinfo{pages}{308--313}
  (\bibinfo{year}{1965}).

\bibitem{cohen2017emnist}
\bibinfo{author}{Cohen, G.}, \bibinfo{author}{Afshar, S.},
  \bibinfo{author}{Tapson, J.} \& \bibinfo{author}{van Schaik, A.}
\newblock \bibinfo{title}{{EMNIST}: extending {MNIST} to handwritten letters}.
\newblock \emph{\bibinfo{journal}{Proc. 2017 International Joint Conference on
  Neural Networks (IJCNN)}} \bibinfo{pages}{2921--2926} (\bibinfo{year}{2017}).

\bibitem{ekec2026supplementary}
\bibinfo{author}{Ekec, K.}, \bibinfo{author}{Kesgin, B.~U.} \&
  \bibinfo{author}{Te{\u{g}}in, U.}
\newblock \bibinfo{title}{Supplementary {Data} for {Single}-shot holographic
  gradients for training wavefronts through disorder}.
\newblock \bibinfo{howpublished}{Dataset} (\bibinfo{year}{2026}).
\newblock \urlprefix\url{https://doi.org/10.5281/zenodo.24286669}.

\bibitem{ekec2026holographicgradients}
\bibinfo{author}{Ekec, K.}, \bibinfo{author}{Kesgin, B.~U.} \&
  \bibinfo{author}{Te{\u{g}}in, U.}
\newblock \bibinfo{title}{{HolographicGradients}}.
\newblock \bibinfo{howpublished}{GitHub repository} (\bibinfo{year}{2026}).
\newblock \urlprefix\url{https://github.com/utegin-lpt/HolographicGradients}.

\end{thebibliography}

\startsupplement

\begin{center}
{\Large \textbf{Supplementary information for}}\\[0.3em]
{\Large \textbf{Single-shot holographic gradients for training wavefronts through disorder}}\\[0.6em]
\textbf{Kerem Ekec$^{1,\dagger}$, Bahad{\i}r Utku Kesgin$^{2,\dagger}$, U\u{g}ur Te\u{g}in$^{1,2,*}$}\\[0.3em]
$^{1}$Department of Biomedical Sciences and Engineering, Ko\c{c} University, \.{I}stanbul 34450, T\"{u}rkiye\\
$^{2}$Department of Electrical and Electronics Engineering, Ko\c{c} University, \.{I}stanbul 34450, T\"{u}rkiye\\[0.3em]
\textit{$^{\dagger}$These authors contributed equally to this work. $^{*}$utegin@ku.edu.tr}
\end{center}

\tableofcontents
\addtocontents{toc}{\protect\setcounter{tocdepth}{2}}

\section{Setup and holographic measurement}\label{sec:setup}

This section describes the optical setup and how one off-axis frame gives the complex object field. \SNote~\ref{sec:optics} gives the optics, the exposures that form one hologram and the encoding of the SLM patterns, \SNote~\ref{sec:demod} the demodulation, its frequency budget and noise, and \SNote~\ref{sec:power} the optical power budget.

\subsection{Optical setup}\label{sec:optics}

A Q-switched laser at 1,030~nm illuminates a phase-only SLM (HOLOEYE PLUTO-2.1 NIR-145), which writes complex amplitudes by the type-2 pixelated encoding of Arriz\'on et al.~\cite{arrizon2007pixelated}; one region of the SLM forms the object arm and another the reference arm. A 4f relay images both; an iris in its Fourier plane passes the first diffraction order only, so a region written without its blazed carrier is switched off. The object beam crosses the diffuser about 53~mm before the camera (52.4--53.9~mm in the fits of the physical model); the reference bypasses it and meets the object beam on the camera at about $3^{\circ}$.

Each hologram records one SLM pattern. Since the pulsed laser cannot be triggered, each exposure (0.1--8~ms) integrates a varying number of pulses, and the energy and the phase between the arms change from exposure to exposure. Unless stated otherwise, each hologram is therefore the mean of five exposures of the same pattern, for the probes of the physical model and the patterns of the measured TM alike, each demodulated, normalized by its energy and aligned in global phase, with outlier exposures rejected by their energy. These exposures carry no information about any other pattern, so measurement costs are counted in SLM patterns throughout; with a stable or triggerable source, one exposure per pattern would suffice.

The object pattern has $32\times32$ SLM cells, each a unit-amplitude phase, written with this encoding, which at unit amplitude is the phase on a blazed grating. The reference pattern $E$ has $100\times100$ complex SLM cells ($|E|\le1$) written in the same way. During object training, the reference region shows a flat phase, that of an untrained plane-wave reference.

\subsection{Off-axis demodulation}\label{sec:demod}

\begin{figure}[!htb]
\centering
\includegraphics[width=\textwidth]{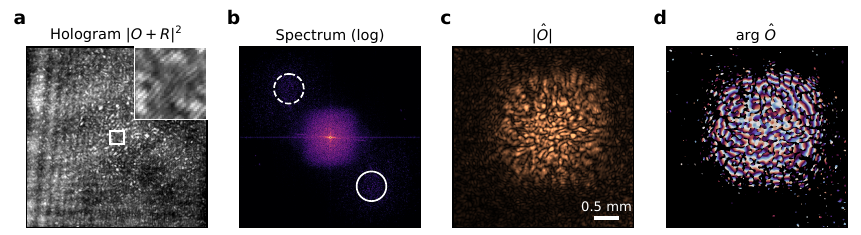}
\caption{\textbf{One off-axis frame gives the complex field.} \textbf{a}, Raw camera frame of a random object pattern through the diffuser with a flat reference; inset, the fringes. \textbf{b}, Its Fourier spectrum; the object sideband (solid circle) and its twin (dashed) sit around $-\mathbf f_0$ and $+\mathbf f_0$, clear of the central terms $|O|^{2}+|R|^{2}$. \textbf{c},\textbf{d}, Amplitude and phase of the demodulated field $\hat O$; phase shown where the amplitude exceeds 10\% of its maximum.}
\label{fig:s_holo}
\end{figure}

A hologram contains $I=|O|^{2}+|R|^{2}+OR^{*}+O^{*}R$. The term $OR^{*}$ sits at the carrier frequency $-\mathbf f_0$ in the Fourier plane ($O^{*}R$ at $+\mathbf f_0$); we filter a disk around $-\mathbf f_0$, shift it to the origin and transform back~\cite{cuche2000spatial}. Dividing by the reference envelope, measured with the object region off, gives $\hat O$ from one frame (\SFig~\ref{fig:s_holo}).

A trained reference is not a plane wave. With spectral half-widths $r_O$ and $r_R=\kappa r_O$ ($\kappa\le1$), separating the sideband from $|O|^{2}$ (radius $2r_O$) and keeping it below Nyquist with the carrier on the diagonal requires $|\mathbf f_0|\ge(3+\kappa)r_O$ and $|\mathbf f_0|/\sqrt2+r_O+r_R\le1/(2L_{\mathrm{pix}})$, so
\begin{equation}
N_{\mathrm{modes}}\approx\frac{\pi N_{\mathrm{pix}}}{4\,\mathcal D(\kappa)^{2}},\qquad \mathcal D(\kappa)=\frac{3+\kappa}{\sqrt2}+1+\kappa ,
\label{eq:s_modes}
\end{equation}
i.e.\ 12.4 camera pixels per mode at $\kappa=0$ (Loterie's condition~\cite{loterie2017microscopy}), 16.0 at $\kappa=1/4$ and 29.7 at $\kappa=1$. Fields are acquired with a flat reference ($\kappa=0$); a structured reference is used only at inference, where the sideband is read without dividing by $R^{*}$ (\SNote~\ref{sec:game}).

In the shot-noise limit the quadrature of $\hat O$ along the reference phase has variance $\approx f_{\mathrm{sb}}(|O|^{2}+|R|^{2})/(2|R|^{2})$ photoelectrons, with $f_{\mathrm{sb}}$ the fraction of the spectrum kept; a reference comparable to the object costs a factor of two in signal-to-noise ratio. Since the phase between the arms changes between exposures, each frame carries an unknown complex factor; it is removed with the model's forward prediction (object arm, \SNote~\ref{sec:grad}) or with a flat-reference pilot read (reference arm, \SNote~\ref{sec:refmodel}). The interferometer therefore needs to be stable only within one exposure.

With all object cells flat, a pattern-independent field remains, coherent with the reference. Its mean, recorded at the start of each session, is subtracted from every field used to fit or score a model; training uses the fields as measured. Two fields $a,b$ are compared by the gauge-invariant correlation $|\langle a,b\rangle|/(\|a\|\|b\|)$; its noise limit is set by the correlation between repeated recordings of one pattern. Throughout, the model accuracy is this correlation between predicted and measured fields on held-out patterns never used for fitting.

\subsection{Optical power budget}\label{sec:power}

An average power of 6~mW illuminates the SLM, whose $1920\times1080$ pixels of 8~$\mu$m pitch span $15.36\times8.64$~mm, or 4.5~mW\,cm$^{-2}$ for uniform illumination. We take the complex-amplitude encoding to send 20\% of this light over the full grid, 1.2~mW, into the first diffraction order passed by the iris, and scale it by the area of each region. Each cell spans $5\times5$ pixels (40~$\mu$m), so the object region ($32\times32$ cells, $1.28\times1.28$~mm) covers 1.23\% of the grid and the reference region ($100\times100$ cells, $4.0\times4.0$~mm) 12.1\%. For unit-amplitude patterns, the object arm therefore carries about 15~$\mu$W, the reference arm about 145~$\mu$W and every cell 14.5~nW. Since the encoding sets the first-order amplitude, a trained reference carries $\langle|E|^{2}\rangle\times145$~$\mu$W, with the mean taken over its cells.

A neutral-density filter (OD~1.0) in the reference arm attenuates it tenfold, which offsets the tenfold larger area of the reference region and brings the two arms to comparable power before the diffuser. The diffuser then scatters much of the object light outside the sensor of the camera (FLIR Blackfly S BFS-U3-04S2M, 12-bit, 6.9~$\mu$m pixels), so the reference stays the stronger arm on the camera, as the off-axis readout favours (\SNote~\ref{sec:demod}). The exposure is chosen so that the frames stay below saturation, and the signal-to-noise ratio of $\hat O$ is then set by the camera and by the pulse-to-pulse jitter of the laser.

\section{Holographic gradients}\label{sec:theory}

This section derives the gradients of both arms from one measured hologram (\SNote~\ref{sec:grad}) and shows why the reference is necessary for them (\SNote~\ref{sec:necessity}).

\subsection{Gradients of both arms}\label{sec:grad}

Camera fields are $O$ (object) and $R$ (reference), $I=|O+R|^{2}$ and $\hat O$ is the measured object field. As in Eq.~(1) of the main text, the object pattern is $X_O=x\odot e^{j\theta}$ and the reference pattern is $E$ ($|E|\le1$). The two SLM regions are lit by real illuminations $A_O$ and $A_R$, so that $O=T_O\{A_O\odot X_O\}$ and $R=T_R\{A_R\odot E\}$; the main text absorbs $A$ into $T$. $\odot$ is the elementwise product, $^{*}$ the conjugate and $^{\mathrm H}$ the adjoint, $\langle Tu,v\rangle=\langle u,T^{\mathrm H}v\rangle$ with $\langle a,b\rangle=\sum_p a_p^{*}b_p$. For a real loss $L$ of $z=u+jv$ we use Wirtinger derivatives~\cite{wirtinger1927formalen},
\begin{equation}
\frac{\partial L}{\partial z^{*}}=\frac12\Big(\frac{\partial L}{\partial u}+j\frac{\partial L}{\partial v}\Big),\qquad \delta L=2\,\mathrm{Re}\Big\langle\frac{\partial L}{\partial z^{*}},\delta z\Big\rangle .
\label{eq:s_dl}
\end{equation}
Since the arms meet only at the camera, the object field does not depend on the reference pattern,
\begin{equation}
\partial O/\partial E=\partial O/\partial E^{*}=0,
\label{eq:s_topology}
\end{equation}
whatever $T_O$ is, provided the two SLM regions do not cross-talk.

The camera records $I=|O+R|^{2}$, and a real loss $L$ depends on the two fields only through $I$. Since $O$ and $R$ do not depend on each other, $\partial I_b/\partial R_b^{*}=O_b+R_b$, and the chain rule with the real $\gamma_b=\partial L/\partial I_b$ gives the same error field for both arms,
\begin{equation}
W_O=\frac{\partial L}{\partial O^{*}}=\gamma\odot(O+R)=\frac{\partial L}{\partial R^{*}}=W_R,\qquad \gamma=\frac{\partial L}{\partial I}.
\label{eq:s_square}
\end{equation}
Each arm maps its SLM pattern $U$ linearly to the camera, $Y=T\{A\odot U\}$ with real $A$. The field $Y$ is then holomorphic in $U$, and
\begin{equation}
\delta L=2\,\mathrm{Re}\Big\langle\frac{\partial L}{\partial Y^{*}},T\,\mathrm{diag}(A)\,\delta U\Big\rangle=2\,\mathrm{Re}\Big\langle\mathrm{diag}(A)\,T^{\mathrm H}\frac{\partial L}{\partial Y^{*}},\delta U\Big\rangle ,
\label{eq:s_linear}
\end{equation}
so the gradient with respect to the pattern is the adjoint of the arm applied to the error field. For the two arms,
\begin{equation}
\frac{\partial L}{\partial E^{*}}=A_R\odot T_R^{\mathrm H}\{W_R\},\qquad \frac{\partial L}{\partial X_O^{*}}=A_O\odot T_O^{\mathrm H}\{W_O\}.
\label{eq:s_both}
\end{equation}
The reference gradient contains no model of the object arm and holds for any object arm that obeys Eq.~(\ref{eq:s_topology}), whereas the object gradient requires $T_O$ to be linear.

For the phase-only object pattern $X_O=x\odot e^{j\theta}$, $\delta X_O=jX_O\,\delta\theta$ gives $\partial L/\partial\theta=2\,\mathrm{Im}[X_O^{*}\odot\partial L/\partial X_O^{*}]=2\,\mathrm{Im}[X_O^{*}\odot A_O\odot T_O^{\mathrm H}\{W_O\}]$, which is Eq.~(3) of the main text. A gradient with respect to the phase of the reference would vanish where $|E|=0$, so we optimize the complex reference $E$ directly on the convex set $|E|\le1$.

The error field depends on the readout. For the object arm, the loss acts on the demodulated intensity, whose cross terms with $R$ are removed by the filter, so $W_O=g\odot\hat O$ with $g=\partial L/\partial|\hat O|^{2}$ (Eq.~(2) of the main text). For the reference arm, the scores of Eq.~(7) of the main text, $s_c=\sum_b M_{c,b}\,\mathrm{Re}(e^{-j\phi}\hat O_bR_b^{*})$, give, with $\mathrm{Re}(z)=\tfrac12(z+z^{*})$,
\begin{equation}
W_R=\frac{\partial L}{\partial R^{*}}=\tfrac12\,e^{-j\phi}\,\hat O\odot\sum_c g_cM_c,\qquad g_c=\partial L/\partial s_c ,
\label{eq:s_sideband}
\end{equation}
The phase $\phi$ between the arms is removed with the pilot read (\SNote~\ref{sec:refmodel}), which gives the form $W_R=\tfrac12\,\hat O\odot\sum_cg_cM_c$ of the main text. Both $W_O$ and $W_R$ are formed from the measured $\hat O$; only $T^{\mathrm H}$ is modelled.

The two arms also differ in the shape of the loss. Consider features $f_b^{(n)}=|u_b^{(n)}+R_b|^{2}$ with $u^{(n)}=T_O\{A_O\odot X_O^{(n)}\}$, centred over the data and entering a convex loss through fixed linear weights. Expanding the square,
\begin{equation}
|u+R|^{2}=|u|^{2}+2\,\mathrm{Re}(u^{*}R)+|R|^{2}.
\label{eq:s_expand}
\end{equation}
As a function of $E$, the first term is constant, the second is linear and the third is common to all data and removed by centring, so the loss is convex in the reference. As a function of $X_O$, $|u|^{2}$ is quadratic and data dependent, so the loss is in general quartic in the object pattern. For the sideband scores $s_c$, which are linear in $R$, the loss is convex in $E$ even without centring. This asymmetry holds even for an exactly known $T_O$, and favours alternating convex steps in $E$ with gradient steps in $\theta$ (\SNote~\ref{sec:joint}).

Each frame returns $\hat O_n=\zeta_nO_n$ with an unknown complex factor (phase between the arms and energy). For the object arm, $\zeta_n=\langle U_n,\hat O_n\rangle/\|U_n\|^{2}$, with $U_n=T_O\{A_O\odot X_{O,n}\}$ the model's forward prediction for the pattern $X_{O,n}=x_n\odot e^{j\theta}$, rotates $\hat O_n$ into the model gauge, and
\begin{equation}
\frac{\partial L}{\partial\theta}=\sum_n 2\,\mathrm{Im}\Big[X_{O,n}^{*}\odot A_O\odot T_O^{\mathrm H}\big\{\bar\zeta_n\,W_{O,n}\big\}\Big],\qquad W_{O,n}=g_n\odot\hat O_n,\qquad \bar\zeta_n=\zeta_n^{*}/|\zeta_n| ,
\label{eq:s_gauge}
\end{equation}
summed over a batch; the losses are energy-normalized, so only this one phase per frame comes from the model. All expressions were checked against finite differences on random instances (relative error $<10^{-8}$).

In practice, the object-arm adjoint is the conjugate transpose of the model's window matrix (focusing, imaging) or automatic differentiation of $2\,\mathrm{Re}\langle\bar\zeta_nW_{O,n},T_O\{A_O\odot X_{O,n}(\theta)\}\rangle$ with $W_{O,n}$ held fixed (classification). Phases start at $\theta=0$ and are updated with Adam; unless stated, the learning rate decays from 0.1 to 0.03.

\subsection{Why a reference is necessary}\label{sec:necessity}

Without a reference, the camera records only $I=|O|^{2}$, and this intensity does not determine the gradient. Let $T_O$ be linear and change the phase of the output field, $O\mapsto O\odot e^{j\psi}$ with real $\psi$. The intensity, the loss and $g$ are unchanged, but the backward field is not,
\begin{equation}
T_O^{\mathrm H}\{g\odot O\odot e^{j\psi}\}\neq T_O^{\mathrm H}\{g\odot O\}
\label{eq:s_necessity}
\end{equation}
in general. Flipping the sign of a single pixel $b$ with $g_bO_b\neq0$, for example, changes the gradient by $-2g_bO_b\,T_O^{\mathrm H}\{\mathbf e_b\}\neq0$. No function of the intensity alone can therefore equal the gradient.

The effect is large. Over 100 random $\psi$, the two gradients are a median $90^{\circ}$ apart (simulation; main Fig.~1c). Physical training schemes therefore measure the cross term by interference~\cite{hughes2018training,xue2024fully,mididoddi2025threading,zhou2026photonic}, while physics-aware training~\cite{wright2022deep} and phase diversity supply the phase from a model or from several frames.

\begin{figure}[!htb]
\centering
\includegraphics[width=0.56\textwidth]{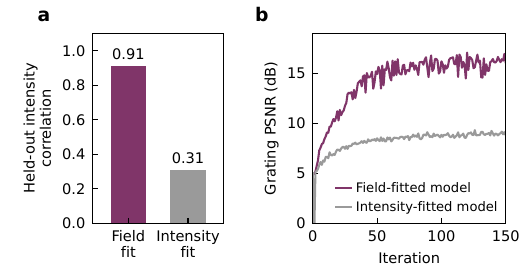}
\caption{\textbf{The reference is necessary.} \textbf{a}, Held-out intensity correlation of a model fitted to measured fields and of one fitted to intensities only; 5,000 probes each. \textbf{b}, Grating PSNR during in-situ training driven by each model.}
\label{fig:s_phase}
\end{figure}

On the bench, 5,000 random probes through the thin diffuser were fitted twice with the same linear model, once to the measured fields and once to the intensities alone by phase retrieval (\SFig~\ref{fig:s_phase}). On held-out probes, the field-fitted model predicts intensities with a correlation of 0.913, the intensity-fitted model with only 0.307. Driving the same in-situ grating loop, the field-fitted model reaches 16.1~dB and the intensity-fitted model 8.9~dB in the last iteration (\SFig~\ref{fig:s_phase}b); in the ten reads after training, the two reach 11.0 and 8.9~dB.
The comparison uses equal probe numbers but not equal noise, since phase retrieval is non-convex. On the bench, loops driven by the physical model and by a measured TM are compared in \SNote~\ref{sec:tm}.

\section{In situ training of the object arm}\label{sec:object}

This section describes the physical model of the object arm (\SNote~\ref{sec:objmodel}), the focusing, image-formation and classification experiments trained through it (\SNote~\ref{sec:objtasks}) and their comparison with a measured TM (\SNote~\ref{sec:tm}).

\subsection{Physical model of the object arm}\label{sec:objmodel}

The fixed free-space path $F$ carries the SLM cells through the relay to the camera plane. Along this path, the cell spectrum is multiplied by the cell aperture, by free-space propagation and by a pupil, the iris with low-order Zernike aberrations in amplitude and phase. The field is then propagated back over the screen distance $z_s$ by the angular-spectrum method~\cite{matsushima2009band}, multiplied by the band-limited screen $D$ and propagated forward to the camera. As in Eq.~(4) of the main text,\begin{equation}T_O=P_{z_s}\,\mathrm{diag}(D)\,P_{-z_s}\,F,\label{eq:s_model}\end{equation}and, since $P_z^{\mathrm H}=P_{-z}$,
\begin{equation}
T_O^{\mathrm H}=F^{\mathrm H}P_{z_s}\,\mathrm{diag}(D^{*})\,P_{-z_s}.
\label{eq:s_adjoint}
\end{equation}

The screen is fitted to the measured probe fields. With $V_k=P_{-z_s}F\{X_k\}$, the measured field is linear in $D$,
\begin{equation}
\hat D=\arg\min_D\sum_k\ell\big(\hat O_k,\;P_{z_s}(V_k\odot D)\big),\qquad \ell(a,b)=1-\frac{|\langle a,b\rangle|^{2}}{\|a\|^{2}\|b\|^{2}},
\label{eq:s_fit}
\end{equation}
where the gauge-invariant misfit $\ell$ leaves each frame a free complex scale, and a small ridge term is added. The prediction is linear in $D$, and for fixed frame scales and geometry the squared misfit would be a convex least-squares problem; the ratio in $\ell$ and the jointly fitted tilts, distance and pupil make the full fit non-convex. Warm starts from the stored screen and early stopping on held-out probes make it converge reliably in practice; we minimize with L-BFGS~\cite{liu1989limited} and exact gradients from reverse-mode differentiation. One probe gives about 6,000 complex values against $\sim\!6\times10^{4}$ screen unknowns that are lit, hence about ten probes; more probes also improve conditioning, since the illuminating speckle $V_k$ has exponential intensity statistics~\cite{goodman2020speckle} and a pixel dark in one probe is bright in another.

Parameters set by the optical configuration, the screen distance, tilts and pupil, are calibrated once from 50 random patterns; a further 200 patterns serve only for evaluation. Before each session, with these parameters fixed, the screen is refitted from 10 probes, starting from the stored screen.

\subsection{Focusing, image formation and classification}\label{sec:objtasks}

Besides one pattern per iteration, each session displays one flat pattern (background field), the 10 probes and 40 evaluation patterns never used by the method. The final pattern is read ten times.

For focusing, the target is a disk of one speckle grain at the centre of the field; the loss is $-\langle|\hat O|^{2}\rangle_{\mathrm{disk}}$. Holograms here average ten exposures instead of five, with shorter exposures in the loop to keep the focus below saturation. The enhancement is $\eta=\langle I\rangle_{\mathrm{disk}}/I_0$, with $I_0$ the mean disk intensity over random patterns.

For image formation, targets of $128\times128$ camera pixels (yin-yang line drawing, clock tower, grating) are low-pass filtered to the resolution of the system. The loss is energy-normalized,
\begin{equation}
L=\big\langle(\sigma|\hat O|^{2}-I_{\mathrm t})^{2}\big\rangle,\qquad \sigma=\textstyle\sum I_{\mathrm t}/\sum|\hat O|^{2},
\label{eq:s_imgloss}
\end{equation}
with $I_{\mathrm t}$ the target; training runs for 150 iterations (grating, clock tower and yin-yang). Image quality is $\mathrm{PSNR}=10\log_{10}(1/\mathrm{MSE})$ of the energy-normalized intensity against the target (peak 1). For the ten final reads we give it in two ways. For the mean image, the ten reads are averaged before scoring (main Fig.~2f and main text), and per frame, each read is scored and the ten values are averaged. Averaging removes frame-to-frame noise, so the mean image scores higher. The mean images of the yin-yang reach 17.2 and 18.1~dB (model, TM), against 17.1 and 17.8~dB per frame, and those of the clock tower 16.7 and 17.1~dB, against 15.1 and 14.3~dB.

For classification, 4,000 FashionMNIST images are split into 2,400 training, 800 validation and 800 test images, resized to $32\times32$ values, one per SLM cell, and written as phase on top of $\theta$. Ten detector squares, in rows of 3--4--3, read the demodulated intensity, and the loss is the cross-entropy of their energy fractions. Each training image is read once per epoch, for eight epochs, and the mask with the best validation accuracy is tested once. The untrained mask ($\theta=0$) with the same detectors is the control (17.8\%); the trained mask reaches 75.0\%. A digital logistic regression on the input phases reaches 82.9\%.

\subsection{Comparison with a measured transmission matrix}\label{sec:tm}

Alongside the model loops, the TM was measured from $3N=3{,}072$ random phase patterns. The measured fields carry an unknown factor per pattern, so $T$ is obtained by ridge regression alternated with the phases of these factors, initialized from the physical model. The TM replaces the model in an otherwise identical loop, so only the adjoint differs.

Over the imaging window ($128\times128$ camera pixels), the TM reaches a model accuracy of 0.947 from 3,072 patterns, the model 0.895 from 10 probe patterns. Its additional accuracy hardly changes training, since the two loops focus to $\eta=97.2\pm1.0$ and $99.5\pm1.6$ and form the yin-yang at 17.24 and 18.10~dB (mean image; 17.1 and 17.8~dB per frame). A single-step focus from the leading singular vector of each adjoint gives $\eta=150\pm5$ (model) and $153\pm6$ (TM). In the probe sweep of main Fig.~2b, scored over the whole field, the model accuracy at 10 probes is 0.77 on the bench and 0.985 in the digital twin, close to the plateaus of 0.82 and 0.989 at 500 probes; a TM from 3,550 probes reaches 0.87 and 0.998. On the bench, the gradient of a model fitted to a single probe already agrees with that of this TM with a cosine of 0.71 (0.89 at 10 probes), whereas a TM fitted to the same 10 probes reaches a model accuracy of only 0.04.
A stored TM also ages, and its model accuracy fell from 0.96 to 0.76 within three hours.

\STab~\ref{tab:budget} lists the SLM patterns spent on each adjoint.

\begin{table}[!htbp]
\centering\small
\caption{SLM patterns spent on the adjoint (thin diffuser). The calibration is shared by all sessions of one optical configuration.}\label{tab:budget}
\begin{tabular}{@{}lrr@{}}
\toprule
 & Physical model & Measured TM \\
\midrule
Configuration calibration (once) & 50 & -- \\
Probes per session & 10 & -- \\
TM patterns & -- & 3{,}072 \\
\bottomrule
\end{tabular}
\end{table}

\section{Offline training of the reference arm}\label{sec:reference}

This section describes the model and offline training of the reference arm (\SNote~\ref{sec:refmodel}), the fringe-matched readout and game play (\SNote~\ref{sec:game}), two-key encryption (\SNote~\ref{sec:enc}) and the joint training of both arms (\SNote~\ref{sec:joint}).

\subsection{Reference-arm model and training}\label{sec:refmodel}

Since the object field does not depend on the reference pattern, the camera image of input $n$ for any reference is $I^{(n)}(E)=|\hat O^{(n)}+R(E)|^{2}$, with $\hat O^{(n)}$ measured once (Eq.~(6) of the main text). The loss and its gradient are therefore available offline for every $E$, and training $E$ on $N_{\mathrm d}$ samples for $N_{\mathrm{it}}$ steps costs $2N_{\mathrm d}$ SLM patterns (acquisition and test) instead of $N_{\mathrm d}N_{\mathrm{it}}$ in situ. The medium must stay static between acquisition and test.

The reference field on the camera is
\begin{equation}
R(\mathbf x)=\sum_{\mathbf f}P_R(\mathbf f)\,C(\mathbf f)\sum_{\mathbf u}A_R(\mathbf u)\,E_{\mathbf u}\,e^{-j2\pi\mathbf f\cdot G\mathbf u}\,e^{j2\pi\mathbf f\cdot(\mathbf x+\mathbf t)} ,
\label{eq:s_refmodel}
\end{equation}
with $P_R$ the iris and free-space propagation of the reference path, $C$ the spectrum of one cell, $A_R$ the illumination, $\mathbf u$ the position of an SLM cell, $G$ the map from SLM cells to camera coordinates and $\mathbf t$ a shift, described by a few geometric and aberration parameters. Its adjoint $T_R^{\mathrm H}$ is analytic and was checked against finite differences.

Calibrated on the bench, the model predicts measured reference fields with a model accuracy of 0.85--0.88, against a repeatability of 0.96--0.99 (\SFig~\ref{fig:s_ref}a).

For demodulated reads, a reference-only frame for the current $E$ is subtracted, whereas the fringe-matched read of \SNote~\ref{sec:game} needs no subtraction. A pilot read (fixed object pattern, flat reference) is repeated regularly, and its phase, interpolated in time, is removed. Object fields are recovered from the sideband with the modelled flat reference.

For training, the loss is the class-weighted cross-entropy of the scores plus a total-variation penalty on $E$. The gradient is $\partial L/\partial E^{*}=A_R\odot T_R^{\mathrm H}\{W_R\}$ with $W_R=\tfrac12\sum_n\hat O_n\odot\sum_cg_c^{(n)}M_c$ summed over the batch (Eq.~\ref{eq:s_sideband}). $E$ starts at unity and is optimized with Adam on its real and imaginary parts, projected onto $|E|\le1$ after every step; the epoch with the best validation accuracy predicted by the model is kept.

\subsection{Fringe-matched readout and game play}\label{sec:game}

\begin{figure}[!htb]
\centering
\includegraphics[width=0.6\textwidth]{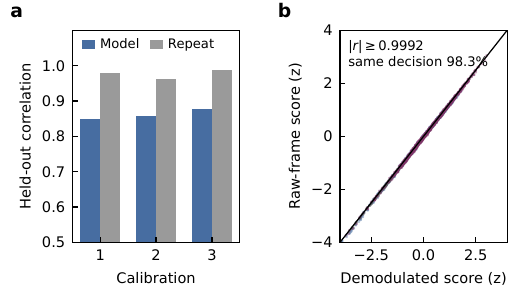}
\caption{\textbf{Reference arm and fringe-matched readout on the bench.} \textbf{a}, Model accuracy (held-out correlation) of the reference-arm model with measured interference, against the repeatability of the measurement, for three calibrations. \textbf{b}, Raw-frame (fringe-matched) versus demodulated scores of the three detectors for the 1,108 Defend the Center test frames, each z-scored.}
\label{fig:s_ref}
\end{figure}

The fringe-matched mask reads the interference term directly. Multiply the raw frame by $M_c(\mathbf x)e^{j2\pi\mathbf f_0\cdot\mathbf x}$ and sum (Eq.~(7) of the main text). The term $OR^{*}$ carries $e^{-j2\pi\mathbf f_0\cdot\mathbf x}$ and becomes slowly varying, so it survives the sum; $|O|^{2}+|R|^{2}$ (bandwidth $\le2r_O<|\mathbf f_0|$) and $O^{*}R$ (at $2\mathbf f_0$) oscillate under the mask and average out, since each detector spans about 17 fringe periods. The read is therefore the sideband score $s_c$ of Eq.~(\ref{eq:s_sideband}) without a Fourier transform; the reference intensity needs no subtraction. On the bench, raw-frame and demodulated scores of the same frames correlate at $|r|\ge0.999$ and give the same decision on 98.3\% of the Defend the Center test frames (\SFig~\ref{fig:s_ref}b) and 99.7\% of the Basic test frames.

ViZDoom~\cite{kempka2016vizdoom} frames map to three actions (left, right, attack), labelled by an oracle. Each scenario has about 10,000 frames, split by episode into training, validation and test sets. Each frame is reduced to one channel by a fixed colour projection, pooled to $32\times32$ values, one per SLM cell, and written as phase; in \emph{Defend the Center}, the maximum of each of the 32 columns over a horizontal band of the frame is written to every cell of that column. Three square detectors form a triangle.

\emph{Two reads.} The scores can be read in two ways. All-optically, the fringe-matched masks above sum the interference term on the raw frame. Alternatively, the sideband is demodulated (\SNote~\ref{sec:demod}), a reference-only frame for the current $E$ is subtracted, and the interference term $\mathrm{Re}(e^{-j\phi}\hat O R^{*})$ is summed over each detector in software; the reference is trained to maximize the classification of these scores in the same way. The demodulated read costs one Fourier transform, filter and inverse transform per frame and one extra reference frame, and the optical inner product is unchanged. We use the demodulated read for \emph{Basic} (main Fig.~3c) and the all-optical read for the harder \emph{Defend the Center} (main Fig.~3d); both reads were computed for both scenarios.

In \emph{Basic}, the demodulated read reaches a test accuracy of 98.7\% (main Fig.~3c); a digital CNN reaches 99.9\% on the same test frames. In a separate run read all-optically, the bench realizes the three detector offsets slightly differently from the model; the raw-frame read reaches 91.8\%, and 94.2\% when these measured offsets are corrected. With this corrected all-optical read, in closed loop on 100 unseen seeds, 100 of 100 episodes are won in $6.88$ steps against $5.85$ for the oracle and the CNN (paired difference $+1.03\pm0.27$).

In \emph{Defend the Center}, every read uses the trained reference as written to the SLM, read all-optically with the fringe-matched masks on the raw frame, without offset correction; demodulated reads of the same frames are given for comparison. On the 1,108 test frames (class shares 45/39/16\%), the raw-frame read reaches a test accuracy of 71.8\% (balanced accuracy 0.738) and the demodulated read 71.3\% (0.734), against 73.6\% predicted by the model. In closed loop on 30 unseen seeds, the raw-frame read scores $7.87\pm0.93$ kills per episode, against CNN $10.67\pm0.63$, linear $3.50\pm0.70$, random $1.37\pm0.20$ and oracle $12.03\pm0.74$; paired on the same seeds, it trails the CNN by $2.80\pm0.71$ and beats the linear classifier by $4.37\pm0.92$ kills.

\subsection{Two-key optical encryption}\label{sec:enc}

\begin{figure}[!htb]
\centering
\includegraphics[width=\textwidth]{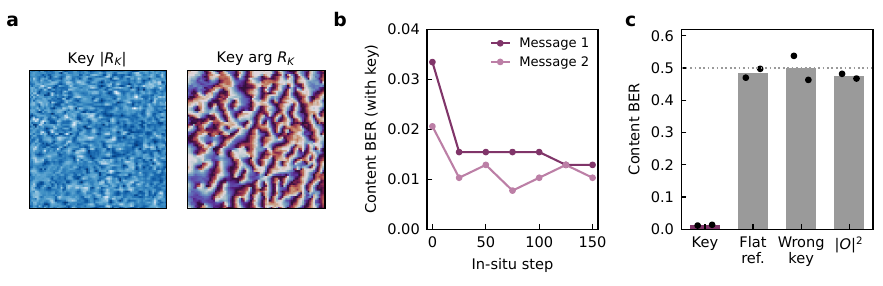}
\caption{\textbf{Two-key encryption on the bench.} \textbf{a}, Amplitude and phase of the realized key field $R_K$ on the code area. \textbf{b}, Content bit error rate of check reads with the key during in-situ optimization of the ciphertext of each message (messages 1 and 2 in the order of the text). \textbf{c}, Content bit error rate with the key (mean of ten decryptions) and under the three attacks; dots, the two messages; dotted line, chance.}
\label{fig:s_enc}
\end{figure}

A QR code (version 2, $25\times25$ modules) is stored in the signs $b_m=\pm1$ of the module sums $z_m=\sum_{\mathbf x\in m}O(\mathbf x)R_K^{*}(\mathbf x)$; finder-pattern modules serve as the phase reference and fixed modules are restored from the standard before decoding. The key $E_K$, a phase-only reference pattern, is optimized through the reference model so that $|R_K|$ is uniform over the code while its phase stays random; the realized key field is then measured from paired reads and used thereafter (\SFig~\ref{fig:s_enc}a). A wrong key is made in the same way from another seed. The ciphertext $\theta_c$ minimizes
\begin{equation}
L=\big\langle\beta^{-1}\mathrm{softplus}[\beta(\mu-u_m)]\big\rangle,\qquad u_m=\frac{b_m\mathrm{Re}(z_me^{-j\alpha})}{\mathrm{rms}|z|},\qquad \alpha=\arg\sum_mb_mz_m,
\label{eq:s_cipher}
\end{equation}
($\beta=4$, $\mu=0.5$), first offline through the object model, then in situ with one flat-reference read and the object-arm gradient per step; each message is carried by one ciphertext frame. The object model is a thin-screen model of the form of \SNote~\ref{sec:objmodel}, fitted from 10 random probes (model accuracy 0.60). Decryption takes one read with $\theta_c$ and $E_K$, sums each module, takes the global phase from the reference modules, thresholds the signs and decodes the QR code with OpenCV. As attacks, we read with a flat reference and with the wrong key, and threshold $|O|^{2}$ with threshold and polarity fitted on the reference modules.

We encrypted the messages ``LABORATORY OF PHOTONICS TECHNOLOGIES'' and ``HOLOGRAPHIC GRADIENTS'' (\SFig~\ref{fig:s_enc}). With the key, the in-situ steps lower the content bit error rate from 0.039/0.023 offline to 0.011/0.014, and all ten decryptions of each message decode. Without the key, the content error stays at chance (0.46--0.54) and nothing decodes.

\subsection{Joint training of both arms}\label{sec:joint}

The system has a trainable input phase $\theta$ ($32\times32$ SLM cells), the diffuser as a fixed mixing layer and a trainable complex reference $E$ ($100\times100$ SLM cells) read by the detectors of \SNote~\ref{sec:game}. Each epoch alternates three steps. (i) $\theta$ is trained in situ. Each training frame of the game is shown with the current $\theta$ and a flat reference and read once, and the class-weighted cross-entropy is back-propagated through the object model. (ii) The object model is refitted on fields of the epoch, kept only if its held-out score improves, and $E$ is trained offline on the new fields. (iii) Validation frames are read with the flat and the trained reference, which gives both the realized reference and the all-optical accuracy. The epoch with the best balanced validation accuracy is kept.

\begin{figure}[!htb]
\centering
\includegraphics[width=0.5\textwidth]{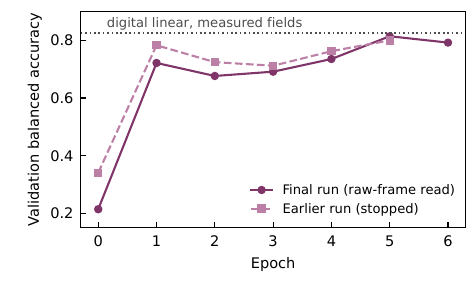}
\caption{\textbf{Joint training on Defend the Center on the bench.} Balanced validation accuracy per epoch for the final run and an earlier run stopped in epoch 6; dotted, digital linear classifier on measured fields.}
\label{fig:s_joint}
\end{figure}

With the ten-probe object model and the raw-frame read (\SFig~\ref{fig:s_joint}), validation balanced accuracy rose from 0.21 to 0.81; on the test set the bench reached 0.746 accuracy and 0.763 balanced accuracy (demodulated read 0.758/0.778) and scored $8.83\pm0.93$ kills in closed loop. A logistic classifier on measured flat-reference fields reaches 0.824 balanced accuracy.

\section{Drifting and thick media}\label{sec:media}

This section describes training while the diffuser drifts (\SNote~\ref{sec:drift}) and through a thick diffuser described by two screens (\SNote~\ref{sec:thick}).

\subsection{Training under drift}\label{sec:drift}

\begin{figure}[!htb]
\centering
\includegraphics[width=0.76\textwidth]{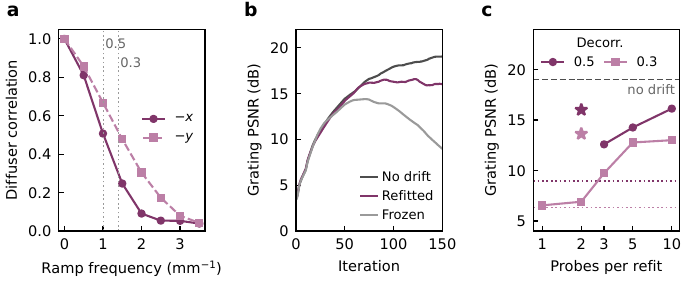}
\caption{\textbf{Training under drift through a thin diffuser on the bench.} \textbf{a}, Correlation of the medium seen by the beam with its initial state versus ramp frequency along two axes, after removal of the output tilt and shift; dotted, final frequencies for decorrelation 0.5 and 0.3. \textbf{b}, Grating PSNR during training (running mean of 10 iterations) without drift; with drift to decorrelation 0.5 and a frozen model, fitted before training and never refitted; and with the same drift and refits from two probes plus the training frame every five iterations. \textbf{c}, Grating PSNR (last 10 iterations) with drift to decorrelation 0.5 versus probes per refit; stars, two probes plus the training frame every five iterations; dotted, frozen model under the same drift; dashed, no drift.}
\label{fig:s_drift}
\end{figure}

The drift experiments used a thin diffuser. Drift is induced by adding to the object pattern a phase ramp along one SLM axis whose frequency grows linearly with the iteration and is hidden from the model. The ramp walks the illumination across the diffuser but also tilts and shifts the output; this tilt and shift were calibrated beforehand and removed from every measured field, so that only the change of the medium seen by the beam remains (\SFig~\ref{fig:s_drift}a). By the end of training, the medium seen by the beam has decorrelated to 0.5 or 0.3.

Each refit repeats the fit of Eq.~(\ref{eq:s_fit}) for the screen and tilts, warm-started, on the most recent probes; training frames are measured fields and enter at no cost. For the grating (main Fig.~4), two new random probes and the latest training frame are added every five iterations.

In focusing, the mean $\eta$ over the last 60 iterations, relative to the drift-free run, is 48\% and 17\% for the frozen model at decorrelation 0.5 and 0.3, and 97\% and 85\% for refits. For the grating at decorrelation 0.5, the PSNR reached with a frozen model falls as the medium drifts, whereas refits keep it close to the drift-free run (\SFig~\ref{fig:s_drift}b). More probes per refit recover more of the drift-free PSNR (\SFig~\ref{fig:s_drift}c); at decorrelation 0.5, ten probes every ten iterations give 16.1~dB against 9.0~dB frozen, and two probes plus the training frame every five iterations give 16.0~dB from fewer probe patterns. These values are the mean per-frame PSNR of the last 10 iterations; main Fig.~4 gives the PSNR of their mean image, 19.5, 9.1 and 17.6~dB without drift, with the frozen model and with refits. Each condition was run once.

\subsection{Memory effect and the thick medium}\label{sec:thick}

Two glass slides, each coated with parafilm on one face, about 15~mm apart, replace the diffuser. A thin screen describes a medium only within its memory effect~\cite{freund1988memory,osnabrugge2017generalized}, within which a tilt of the input tilts the output speckle without changing it. We measure this by tilting a plane-wave input, registering the output tilt and correlating each output with the untilted one.

\begin{figure}[!htb]
\centering
\includegraphics[width=\textwidth]{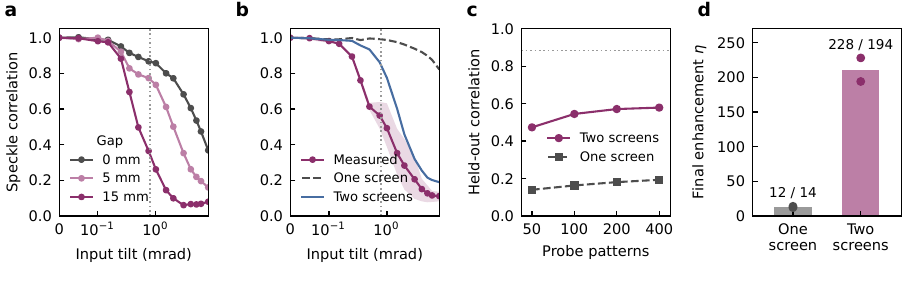}
\caption{\textbf{On the bench, a thick medium needs two screens.} \textbf{a}, Speckle correlation versus input tilt for two parafilm-coated slides in contact and 5 or 15~mm apart (mean of $\pm x$, $\pm y$). Dotted, angular spacing of the input modes $\lambda/w_O$. \textbf{b}, The stack of main Fig.~4, measured (band, range over the four directions) and predicted by a single-sheet model and by the fitted two-screen model. \textbf{c}, Model accuracy of one- and two-screen models versus probes; two screens, the sweep of main Fig.~4h; one screen, an earlier fit of the stack; dotted, repeatability ceiling of the sweep (0.88). \textbf{d}, Final enhancement of focusing trained through each model (two runs each).}
\label{fig:s_thick}
\end{figure}

The memory effect collapses with the gap. With the slides in contact the speckle correlation falls to 0.5 only at 5.7~mrad, but with the slides 15~mm apart already at 0.5~mrad (\SFig~\ref{fig:s_thick}a). For the stack of main Fig.~4, it falls to 0.5 at 1.0~mrad, comparable to the angular spacing of the input modes, $\lambda/w_O=0.80$~mrad with $w_O=1.28$~mm the width of the object pattern (\SFig~\ref{fig:s_thick}b). A single-sheet model cannot describe this decay; the fitted two-screen model reproduces its shape, decaying about twice as slowly. The gap scan used an earlier mounting of the stack.

Two screens describe the stack far better than one. From 400 probes, two screens reach a model accuracy of 0.58, whereas one screen reached only 0.19 in an earlier fit of the stack (\SFig~\ref{fig:s_thick}c). In that fit, two screens with free distances converged to a gap of 14.8~mm, matching the 15~mm measured. Training through each of these models (200 fit probes) focuses to $\eta=12$ and 14 with one screen and 228 and 194 with two (\SFig~\ref{fig:s_thick}d). The two focusing runs of main Fig.~4f, trained through the two-screen model of the sweep, reach 207 and 193.

For main Fig.~4e--h, the two screens are combined with the input model of \SNote~\ref{sec:objmodel}. The sweep of main Fig.~4h fits the model from up to 400 probes; the training runs use the model fitted from 200 probes.
Focusing and image formation (grating, EMNIST letters~\cite{cohen2017emnist}) run for 150 iterations each.

\begin{figure}[!htb]
\centering
\includegraphics[width=\textwidth]{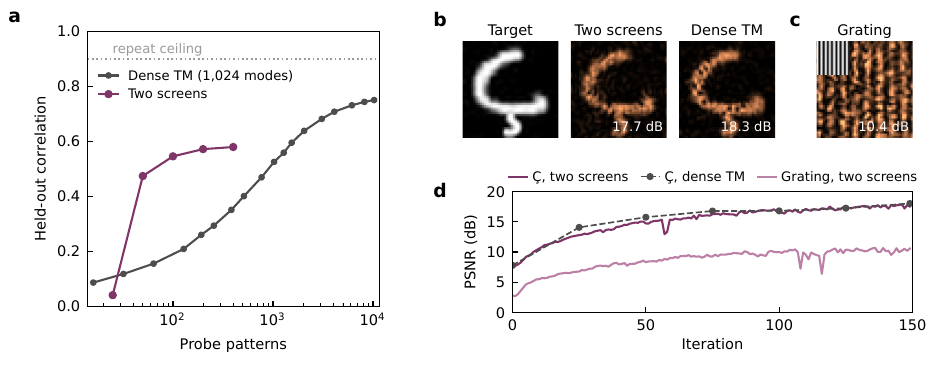}
\caption{\textbf{Limits in the thick medium on the bench.} \textbf{a}, Model accuracy of a dense TM (1,024 modes) and of the two-screen model versus probes; dotted, repeatability ceiling of the TM run (0.90). \textbf{b}, Ç formed through the two-screen model and through the dense TM (PSNR of the mean of ten reads). \textbf{c}, Grating formed through the two-screen model; inset, target. \textbf{d}, PSNR during training through the thick diffuser (read of five exposures per iteration); dense TM, logged every 25 iterations. Intensities normalized to the target energy.}
\label{fig:s_limits}
\end{figure}

The two-screen fit is limited by its model class rather than by probes. Beyond 200 probes, its model accuracy (0.57--0.58) approaches its training accuracy and stays below the repeatability ceiling of the sweep, 0.88. A dense TM of all 1,024 modes, measured from 10,240 patterns with one exposure each, reaches 0.75 and is still rising, against a repeatability ceiling of 0.90 for its own single-exposure frames; it passes the two-screen model at about 1,500 probes (\SFig~\ref{fig:s_limits}a). The stack also has a fast fluctuating part, about 17\% of the field energy, which caps any static model. Forming the letter Ç, the dense TM reaches 18.3~dB and the two-screen model 17.7~dB (\SFig~\ref{fig:s_limits}b), and both rise at a similar rate during training (\SFig~\ref{fig:s_limits}d). A grating formed with two screens reaches 10.4~dB (\SFig~\ref{fig:s_limits}c).

\section{Scaling with the number of modes}\label{sec:scaling}

\begin{figure}[!htb]
\centering
\includegraphics[width=\textwidth]{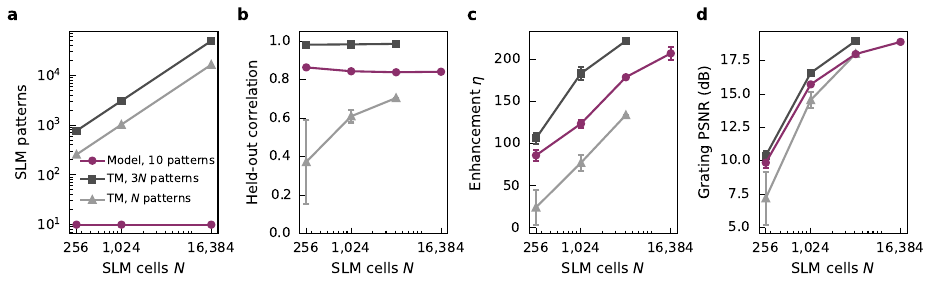}
\caption{\textbf{Scaling with the number of modes in the digital twin.} \textbf{a}, SLM patterns per adjoint. \textbf{b}, Model accuracy. \textbf{c}, Enhancement after 150 iterations. \textbf{d}, Grating PSNR after 150 iterations. Mean $\pm$ s.d. over random diffusers. No TM at $N=128^{2}$, which requires more memory.}\label{fig:s_scaling}
\end{figure}

The cost of a TM grows with the number $N$ of controlled modes, since it needs at least one pattern per SLM cell, whereas the physical model of the object arm has a size set by the optics, since the screen is band-limited by the relay and the iris, and each probe supplies as many complex values as the demodulation band holds, independently of $N$. The number of probes needed to fit the screen should therefore stay nearly constant as the system grows. To test this, we repeat the focusing and image-formation experiments in a digital twin of the setup for $N=16^{2}$, $32^{2}$, $64^{2}$ and $128^{2}$ SLM cells at a fixed number of probes. The twin propagates the SLM field through a random thin screen and an aberrated pupil to the camera, adds the noise of the bench (wavefront and energy jitter of the laser, shot and read noise, camera quantization) and demodulates off-axis frames as on the bench; the screen and the aberrations are hidden from the methods. The cell size, iris, distances, light level per cell and noise are the same at every $N$, and only the aperture and the camera chip grow. Since the object band is set by the iris, it does not widen with $N$, so a larger $N$ needs a larger camera chip rather than finer pixels. For each $N$, we compare the physical model fitted from $K=10$ probes with a TM measured from $N$ and $3N$ patterns, by their model accuracy, the cosine between their gradient and that of the true system, and the enhancement and grating PSNR reached in situ after 150 iterations (\SFig~\ref{fig:s_scaling}). At $N=128^{2}$, the TM was not computed, since it requires more memory than our workstation has.

The model accuracy of the ten-probe model stays between 0.84 and 0.87 from $N=256$ to 16,384 (\SFig~\ref{fig:s_scaling}b), and the cosine between its gradient and the true gradient, not plotted, between 0.85 and 0.88. At $N=32^{2}$, this model accuracy (0.85) is close to that of the model on the bench (0.90 over the imaging window, \SNote~\ref{sec:tm}). A TM from $N$ patterns, which already costs far more patterns than the model (\SFig~\ref{fig:s_scaling}a), reaches a model accuracy of only 0.37 to 0.71 and trains the system to a lower enhancement than the model at every size (\SFig~\ref{fig:s_scaling}c). A TM from $3N$ patterns is nearly exact in the twin (0.98 to 0.99), more so than on the bench (0.95), since the twin has no slow drift during the acquisition, and reaches 1.2 to 1.5 times the enhancement of the model and a 0.5 to 1.0~dB higher grating PSNR (\SFig~\ref{fig:s_scaling}c,d). With its cost fixed at 10 patterns, the model follows the growth of the TM with $N$, with an enhancement rising from 86 to 208 and a grating PSNR from 9.9 to 18.9~dB, including at $N=128^{2}$, where a TM would need 16,384 to 49,152 patterns. The cost of the model therefore stays constant while that of the TM grows linearly with $N$, so the advantage of the model widens with the size of the system.

\section{Metrics and statistics}\label{sec:metrics}

Images are scored by the PSNR of the energy-normalized intensity against the target (peak 1), on the camera grid. Focusing is scored by $\eta$. Single reads are reported as mean $\pm$ s.e.m.\ over the ten final reads, together with the PSNR of their mean. Classifiers are scored by test accuracy and balanced accuracy (mean recall), and closed-loop policies were compared on identical seeds with paired tests. Encryption is scored by the bit error rate over content modules.

\end{document}